\documentclass[12pt,preprint,times]{elsarticle}
\usepackage{color}
\usepackage{multirow}
\usepackage[ansinew]{inputenc}
\usepackage[psamsfonts]{amssymb}
\usepackage{amsmath}
\usepackage[all]{xy}
\usepackage{graphicx}
\usepackage{float}
\usepackage{xspace}

\usepackage{rotating}

\input{epsf}

\journal{Acta Tropica}
\begin{document}
\begin{frontmatter}
\title{Ross-Macdonald Models: Which one should we use? }
\date{}
\author[address1,address3]{Mario Ignacio Simoy}
\author[address1,address2,email]{Juan Pablo Aparicio}

\address[address1]{Instituto de Investigaciones en Energ\'ia no Convencional (INENCO), Consejo Nacional de Investigaciones Cient\'ificas y T\'ecnicas (CONICET),
Universidad Nacional de Salta, Av. Bolivia 5100, 4400 Salta,
Argentina.}
\address[address2]{Simon A. Levin Mathematical, Computational and Modeling Sciences Center, Arizona State University, PO Box 871904 Tempe, AZ 85287-1904, USA}

\address[address3]{Instituto Multidisciplinario sobre Ecosistemas y Desarrollo Sustentable, Universidad Nacional del Centro de la Provincia de Buenos Aires, Paraje Arroyo Seco s/n, 7000 Tandil, Argentina.}

\address[email]{Corresponding author: juan.p.aparicio@gmail.com}

\begin{abstract}

\noindent
{\color{red}
Ross-Macdonald models are the building blocks of most vector-borne disease models. Even for the same disease, different authors use different model formulations, but a study of the dynamical consequences of assuming different hypotheses is missing. In this work we present different formulations of the basic Ross-Macdonald model together with a careful discussion of the assumptions behind each model. The most general model presented is an agent based model for which arbitrary distributions for latency and infectious periods for both, host and vectors, is considered. At population level we also developed a deterministic Volterra integral equations model for which also arbitrary distributions in the waiting times are included. We compare the model solutions using different distributions for the infectious and latency periods using  statistics, like the epidemic peak, or epidemic final size, to characterize the epidemic curves. The basic reproduction number ($R_0$) for each formulation is computed and compared with empirical estimations obtained with the agent based models. The importance of considering realistic distributions for the latent and infectious periods is highlighted and discussed. We also show that seasonality is a key driver of vector-borne disease dynamics shaping the epidemic curve and its duration.
}

\end{abstract}

\begin{keyword}

Ross-Macdonald Model \sep Epidemiology \sep Delayed Model \sep Agent Based Model

\end{keyword}
\end{frontmatter}

\section{Introduction}

{\color{red}

Vector-borne diseases are caused by different types of parasites, including viruses and bacteria,
which are transmitted by vectors as mosquitoes,
sandflies, ticks, and kissing-bugs, among others. According to the World Health
Organization, every year there are more than 700 thousand
deaths as a consequence of vector-borne diseases \cite{who2017}.

Mosquito-borne diseases of humans include malaria, dengue, zika,
chykungunya, yellow fever (see for example \cite{snow2015, benelli2016, gardner2017}).
Different triatomine species transmit {\it Trypanosoma cruzi} the causal agent of Chagas disease (see for example \cite{schofield1997} and references there in) while Leishmaniasis is  transmitted by several species of sandflies \cite{bates2007}.}

Vector-borne diseases are also common zoonotic diseases. Some
forms of Leishmaniasis cannot be transmitted from humans to
sandflies and the parasite population survives in a wild cycle
including small rodents, {\color{red} dogs, cows and several species of birds as hosts \cite{anaguano2015,bates2007}.} West Nile Virus may be
transmitted to humans but it is maintained in a cycle which
includes several species of birds \cite{wonham2008}. \textit{Trypanosoma cruzi}, the causal
agent of Chagas disease is also transmitted to different
animals including dogs, marsupials, rodents, and others hosts \cite{noireau2009}.

Ross model was published in 1911 \cite{ross1911} and remains as the basis of countless models for vector-borne diseases.
Ross considered a simple model for malaria, with births and deaths but with constant populations and infectious periods exponentially distributed. {\color{red}Humans and mosquitoes may be in only two classes}: Affected and Unaffected (what here we will denoted by  $H_i, H_s, V_i, V_s$). Then, Ross model in continuous time reads

\begin{align}
\frac{dH_i}{dt} &=\beta_h m \frac{V_i}V (H-H_i)-r_h H_i  \nonumber\\ \nonumber
\frac{dV_i}{dt} &=\beta_v \frac{H_i}H  (V-Vi)-\mu_v V_i \\ \nonumber
\end{align}
where {\color{red} $H$ and $V$ are the numbers of humans and mosquitoes}, $m$ is the number of mosquitoes per human ($V/H$), $r_h$ is the recovery rate for humans, $\mu_v$ is the mortality rate for mosquitoes, and $\beta_j$ are the transmission parameters which may be decomposed as $\beta_j=bfp_j$ with $b$ the mosquitoes biting rate, $f$ the proportion of bites in humans, and $p_j$ the probability of transmission per bite. Ross formulation is still used but it is not advisable. The parameter $m$ is in fact a dynamical variable. For the original Ross model this was not a problem as he considered constant populations. However, both vector and host populations may vary in time, therefore, an equivalent formulation, more frequently used, and preferable is

\begin{align}
\frac{dH_i}{dt} &=\beta_h {V_i}\frac {H_s}H -r_h H_i \label{rossmodel1} \\
\frac{dV_i}{dt} &=\beta_v V_s\frac{H_i}H -\mu_v V_i  \label{rossmodel2}
\end{align}

{\color{red} Models with these rates of infection are broadly known as Ross-Macdonald models, albeit  Macdonald's contribution to the Ross model is not reflected in this model formulation. Macdonald modified the original Ross model, integrating biological information about the mosquito latency period, and introduced the exposed class for vectors  \cite{mandal2011}. Later he considered also the case of super-infection in Malaria disease dynamics (see for an extensive discussion \cite{smith2012})}.

For the Ross model, the basic reproduction number ($R_0$), defined as the number of
secondary host cases produced by a typical infectious host in a
completely susceptible population is
\begin{equation}
R_0=\frac{\beta_h\beta_v}{r_h\mu_v}\frac VH
\nonumber
\end{equation}

This celebrated result from Ross \cite{ross1911} shows that the basic
reproduction number is proportional to the number of vectors per
host ($V/H$), {\color{red}and therefore, disease transmission may be interrupted if the number of vectors per host is reduced below some threshold.

Since the pioneering work of Ross, several extensions of his basic
model (Eqs. \ref{rossmodel1} - \ref{rossmodel2}) were developed including the
addition of exposed classes, superinfection, spatiality,
time-varying populations, age structure and more (see for example
\cite{dietz1974,bacaer2007,auger2008,mandal2011,smith2012,oregan2016,sanchez2020}),
and applied to the study of different infectious diseases such as
malaria, dengue, yellow fever, cutaneous Leishmaniasis, Chagas
disease, West Nile virus, among others (see for example
\cite{wilder-smith2018,smith2004,mandal2011,amaku2016,bacaer2006,
velasco1991,wonham2008} and references therein). }

{\color{red} In this work we present a detailed analysis of some general Ross-Macdonald models. We show that the inclusion of exposed classes as well as the distribution of the latent and infectious periods, have significant dynamical consequences.  We also show that seasonality is a major factor shapening the epidemics curves.}

{\color{red}This paper is arranged as follow. In the next section we discuss the general assumptions common of all  models presented. In Section 3  several deterministic Ross-Macdonald models are developed considering exposed classes and different distributions for the waiting periods. The basic reproduction number is computed in each case. A stochastic agent based model (ABM) is developed in Section 4. Numerical results, such as epidemic curves, epidemic final sizes, the basic reproduction number are computed for each model and compared between them in Section 5. The key role of seasonality is also discussed. Finally present the discussion of the results and conclusions.}

\section{General assumptions and parameters}

In a Ross-Macdonald model it is assumed that populations are homogeneously mixed. Vector's bites are evenly divided among hosts, that is, every time a vector bites, chooses a host at random. This hypothesis leads to a frequency dependent transmissions terms proportional to $V_iH_s/H$ and $V_sH_i/H$. {\color{red}This central hypothesis is perhaps what define what a Ross-Macdonald model is. However we want to stress that this assumption is only realistic for small populations like a household. The use of Ross-Macdonald type of models for larger populations will be analyzed elsewhere.}

{\bf Demography}. {\color{red} Immigration and emigration are not considered as we are interested in the simplest cases.} Births are assumed to take place at a (density-independent) rate $\Lambda$. Deaths may be described by the mortality or by the survival function.
Mortality ($\mu$) is the number of deaths per individual and per unit of time. In general it is an age-dependent rate. The survival function, $\bar F(a)$, is the proportion of individuals still alive at age $a$, and it is related with the mortality by $\bar F(a)=1-e^{-\int_0^a\mu(s)ds}$.

{\bf Epidemiology}. Populations are divided in some of the following epidemiological classes: Susceptible, Latent, Infectious, and Recovered.
Latent (or Exposed) individuals are infected but not infectious (and therefore are unable to transmit the disease). Recovered individuals are immune, and therefore do not participate of the transmission process. Duration of the latent period may be described for a survival function of the age of infection, $\bar F_e(s)$ which gives the proportion of latent individuals who remain latent at age of infection $s$ (age of infection is the time elapsed since first infection).
Analogously, $\bar F_i(s)$ is the proportion of infectious individuals who remain infectious after a time $s$ after the end of latency.
Alternatively we can use the, (in general) age-of-infection dependent, progression rates (from latency to infectiousness) or recovery rates (from infectiousness to recovery).

All the periods considered (lifespan, latency period, infectious period) are random variables which may be characterized by a probability distribution. The simple, and commonly used case of exponentially distributed periods correspond to constant, age independent, rates. For example using a constant mortality rate $\mu$ imply the assumption of an exponentially distributed lifespan.

Parameters defining the different periods distributions are:
\begin{description}
\item \hspace{1cm} $T_h$: Host life expectancy (mean lifespan)
\item \hspace{1cm} $T_v$=$T_{vi}$: Vector life expectancy, mean infectious period for vectors
\item \hspace{1cm} $T_{he}$: Mean latency period for exposed hosts
\item \hspace{1cm} $T_{hi}$: Mean infectious period for hosts
\item \hspace{1cm} $T_{ve}$: Mean latency period for vectors
\end{description}
{\color{red}For the limiting cases of exponentially distributed or fixed periods these parameters values completely define the probability distributions. In the general case other parameters like the variance of the distribution should be provided.}
In all cases we considered that vectors are infectious for life.

{\bf Entomological parameters}.  Biting rate on hosts (number of bites per vector, per unit of time, on hosts) is denoted by $b$. Probabilities of transmission per bite are $p_h$ and $p_v$ (from vectors to hosts and from hosts to vectors respectively). Finally we define $\beta_h=p_hb$, and $\beta_v=p_vb$.

{\bf Basic reproduction numbers}. For a general Ross-Macdonald model the basic reproduction number
may be obtained by simple bookkeeping \cite{diekman2000}. One infectious host will produce an average of $\beta_v V\frac 1H$ infected vectors per unit of time. If the mean infectious period for hosts is $T_{hi}$, then the total number of infected vectors is  $\beta_v V\frac 1H T_{hi}$. Only a fraction $f_v$ will survive the latency period, and therefore, the total number of infectious vectors produced by the initial infectious host is $\beta_v V\frac 1H T_{hi}f_v$. Each infectious vector would produce  $\beta_h T_{vi}$ host infections ($T_{vi}$ is the mean infectious period for vectors) and only a fraction $f_h$ will survive the host latency period. Finally the basic reproduction number is given by
\begin{equation}
R_0=\beta_h\beta_vT_{hi}T_{vi}f_hf_v\frac VH
\label{R0}
\end{equation}

\section{Deterministic Ross-Macdonald models}

In a Ross-Macdonald model there are host and vector populations (of size $H$ and $V$ respectively) homogeneously mixed. Each population is subdivided in epidemiological classes. For example, susceptible and infectious host and vector populations ($H_s$, $H_i$, $V_s$ $V_i$). Vectors bite at the rate  $b$ (daily number of bites per vector, for example). If $p_h$ is the probability of infection transmission to hosts per bite, $p_v$ the probability of vector infection per bite on infectious hosts, then, the rate of infection of susceptible hosts is given by $p_hbV_i\frac{H_s}{H}$ while the rate of infection of susceptible vectors by $p_vbV_s\frac{H_i}{H}$. These functional forms for the infection rates are characteristic of all the Ross-Macdonald type models. In the following we will present, discuss and compare the more common deterministic models (without age structure).

\subsection{Basic Model}
One of the most {\color{red} simple, general, and used model is the}  $SIR$ model for hosts and a $SI$ model for vectors. Mortalities are denoted by $\mu$ while recovery rates by $r$. $\Lambda$'s are the  recruitment rates. We will assume that all the periods are exponentially distributed and therefore we obtain the following {\it Basic model}:

\begin{align}
\frac{dH_s}{dt} &=\Lambda_h-\beta_hV_i\frac{H_s}{H}-\mu_hH_s\label{mod1Hs}\\
\frac{dH_i}{dt} &=\beta_hV_i\frac{H_s}{H}-(r_h+\mu_h)H_i\\
\frac{dH_r}{dt} &=r_h H_i-\mu_hH_r\\
\nonumber\\
\frac{dV_s}{dt} &=\Lambda_v-\beta_vV_s\frac{H_i}{H}-\mu_vV_s\label{mod1Vs}\\
\frac{dV_i}{dt} &=\beta_vV_s\frac{H_i}{H}-\mu_vV_i\label{mod1Vi}
\end{align}
where $\mu_h=1/T_h$ and $\mu_v=1/T_v$. Mean infectious period for
host includes recovery and mortality, and therefore in this case
is given by $T_{hi}=1/(r_h+\mu_h)$, from where recovery rate $r_h$
can be estimated. In this work we will consider only the case
$\mu_h=0$, but for many species of hosts, $\mu_h\ll r$ and
therefore we may approximate the recovery rate by $1/T_h$. Vectors
are assumed to be infectious for life and then
$\mu_v=1/T_{vi}=1/T_v$.

Because in this model there are not latency periods, $f_h=f_v=1$ and the basic reproduction number (Eq. \ref{R0})   becomes
\begin{equation}\label{R01}
R_0^{(1)}=\frac{\beta_h\beta_v}{(r_h+\mu_h)\mu_v}\frac{V}{H}
\end{equation}

The assumption of constant mortality for vectors is plausible as for insects we expect an approximately constant daily probability of death. For hosts like birds, constant mortality is also usually observed. However hosts like humans present a survival of type I: low mortality for ages below the mean followed by a steep decrease in survival.
In this case an age structured model for the host population should be used (see for example \cite{sanchez2020}). However in those cases we have that $\mu_h\ll \mu_v$ and therefore we may disregard birth and deaths in the host population when studying the short-term dynamics like in a single outbreak, the case we are considering in this work.

Infectious period is also assumed exponentially distributed, a not realistic assumption. Hosts may lose immunity becoming susceptible again, a case we do not consider in this work.

\subsection{Basic Model with exposed classes}
For both, hosts and vectors, there are latent periods and therefore a more realistic model is a $SEIR$ model for hosts and a $SEI$ model for vectors (as in most cases vectors are infectious for life). The basic model with latent classes (SEIR-SEI model) is:

\begin{align}
\frac{dH_s}{dt} &=\Lambda_h-\beta_hV_i\frac{H_s}{H}-\mu_hH_s\label{modseirHs}\\
\frac{dH_e}{dt} &=\beta_hV_i\frac{H_s}{H}-(k_h+\mu_h)H_e\\
\frac{dH_i}{dt}&=k_hH_e-(r_h+\mu_h)H_i\\
\frac{dH_r}{dt}&=r_hH_i-\mu_hH_r\\
 \nonumber \\
\frac{dV_s}{dt}&=\Lambda_v-\beta_vV_s\frac{H_i}{H}-\mu_vV_s\\
\frac{dV_e}{dt}&=\beta_vV_s\frac{H_i}{H}-(k_v+\mu_v)V_e\\
\frac{dV_i}{dt}&=k_vV_e-\mu_vV_i\label{modseirVi}
\end{align}

Here, $k_h$ and $k_v$ are the progression rates from latency to infectiousness, and in this context are given by $k_j=1/T_{je}$ with $T_{je}$ the mean latency periods ($j=h$ for hosts, and $j=v$ for vectors).

In this case the basic reproduction number is
\begin{equation}
R_0^{(2)}=\frac{\beta_h\beta_v}{(r_h+\mu_h)\mu_v}  \left(
\frac{k_v}{k_v+\mu_v}\right)  \left(\frac{k_h}{k_h+\mu_h}\right)
  \frac{V}{H}\label{R02}
\end{equation}
where $f_j=k_j/(k_j+\mu_j)$ are the fractions of exposed individuals who survive the latency period.

The assumptions in this model are the same discussed above but here it is also assumed that latent periods are exponentially distributed a not realistic assumption neither. Once again $k_h\gg\mu_h$ and then $\frac{k_h}{k_h+\mu_h}\approx 1$.

\subsection{Models with arbitrary distributions for the waiting periods}

The assumption of exponentially distributed periods is appealing because the corresponding ODE models have constant parameters.
However latency or infectious periods are, in general, random variables with non-exponential distributions.

In our case, where we are considering that vectors remain infectious for life, the infectious period is the vector lifespan. In this case a constant mortality is a realistic choice and therefore the infectious period is exponentially distributed. However this is not the case of vector's latent period or the latent and infectious host's periods.

As an example we will first consider the simple case of a $SIR-SI$ model. For vectors we have the equations \ref{mod1Vs}-\ref{mod1Vi}.
For the host population we will consider that the infectious period  ($T_{hi}$) is a random
variable with probability distribution function $f(s)$. As usual,
the cumulative distribution is denoted by $F(s)$. The complementary
cumulative distribution, $\bar{F}(s)=1-F(s)$, is known as the
survival function and gives the probability that an individual
infected in $s=0$ remains infected at time $s$. Because only the fraction $\bar{F}(t-s)$ of the infections produced at time $s$ survives until time $t$ we obtain the integral Volterra equations

\begin{align*}
H_s(t) &=H_s(0)-\int_0^t\frac{\beta_h}{H}V_i(s){H_s(s)}ds  \\
H_i(t) &= H_i(0)\bar{F}(t)+\int_0^t\frac{\beta_h}{H}V_i(s){H_s(s)}\bar{F}(t-s)ds   \\
H_r(t) &= H-H_s(t)-H_i(t)
\end{align*}

Differentiation of Volterra equation gives the following system of integro-differential equations,

\begin{align*}
\frac{dH_s}{dt} &=-\beta_hV_i\frac{H_s}{H}\\
\frac{dH_i}{dt} &=H_i(0)\frac{d\bar{F}}{dt} + \beta_hV_i\frac{H_s}{H}\bar{F}(0)+\int_0^t\frac{\beta_h}{H}V_i(s){H_s(s)}\frac{d\bar{F}}{dt}(t-s)ds \\
&=-H_i(0)f(t) + \beta_hV_i\frac{H_s}{H}-\int_0^t\frac{\beta_h}{H}V_i(s){H_s(s)}f(t-s)ds
\end{align*}

Realistic distributions for infectious or latent periods are bell shaped and therefore survival function is of type I.
Then, a simple but realistic distribution is obtained for the limiting case of fixed infectious period $T_{hi}$. In this case the survival function is a step function, the probability density distribution is
$\delta(t-T_{hi})$, and therefore we obtain the delayed equation

\begin{equation}
\frac{dH_i}{dt} =-H_i(0)\delta(t-T_{hi})+\beta_hV_i\frac{H_s}{H}-\beta_hV_i(t-T_{hi})\frac{H_s(t-T_{hi})}{H} \\
\end{equation}

{\color{red} In the general case of arbitrary distributions in latency and infectious periods for hosts and vectors we have the following Volterra integral equations model,

\begin{align}
H_s(t) &=H_s(0)-\int_0^t \beta_h V_i(s)\frac{H_s(s)}{H}ds \label{volterra1} \\
H_e(t) &= H_e(0)\bar{F}_{he}(t)+\int_0^t \beta_h V_i(s)\frac{H_s(s)}{H} \bar{F}_{he}(t-s)ds   \\
H_i(t) &= H_i(0) \bar{F}_{hi}(t) + \int_0^t \int_0^\tau \beta_h  V_i(s)\frac{H_s(s)}{H}\left[-\frac{d\bar{F}_{he}}{dt}(\tau -s)\right]\bar{F}_{hi}(t-\tau) \; ds d\tau \\
H_r(t) &= H-H_s(t)-H_i(t)-H_e(t) \\ \nonumber
\\
V_s(t) &= V_s(0) e^{-\mu t} + \int_0^t \Lambda_v e^{-\mu(t-s)} ds - \int_0^t \beta_v V_s(s)\frac{H_i(s)}{H}ds \\
V_e(t) &= V_e(0) e^{-\mu t} + \int_0^t \beta_v V_s(s)\frac{H_i(s)}{H} \bar{F}_{ve}(t-s) e^{-\mu(t-s)}ds \\
V_i(t) & = V_i(0) e^{-\mu t} + \int_0^t \int_0^\tau \beta_v V_s(s)\frac{H_i(s)}{H} \left[-\frac{d\bar{F}_{ve}}{dt}(\tau -s)\right] e^{-\mu(t-s)} \; ds d\tau \label{volterra5}
\end{align}
where $\bar F_{je}$, $\bar F_{ji}$ are the survival functions for the exposed and infectious populations  ($j=h$ for host and $j=v$ for vectors).
 }

{\color{red}
\subsubsection{Gamma distributed periods} \label{sec:gamma_dist}

Realistic probability distribution functions for infectious or latent periods are bell shaped and therefore the survival functions are of type I. While accurate numerical solutions of a system of ordinary differential equations like model \ref{modseirHs} - \ref{modseirVi} are easily obtained using a Runge-Kutta scheme, for example,  integral systems like \ref{volterra1}-\ref{volterra5} are not that amenable.

Gamma distributions are flexible functions with two parameters,
the shape parameter $k$ and the scale parameter $\theta$. Some
features of this distribution are particularly appealing. The
exponential distribution is a special case of the Gamma
distribution when $k=1$, while for $k\rightarrow\infty$ the Gamma
distribution converges to the Dirac delta function. Most
importantly, for integer values of $k$ the system
(\ref{volterra1}-\ref{volterra5}) is equivalent to a system of
ordinary differential equations with constant rates (see for example \cite{smith2011} and \ref{sec:linear-trick}).  This result allows to obtain numerical solutions of the
system of integral equations using a simple numerical scheme like
Runge-Kutta.

}

\subsubsection{Delayed Model}

{\color{red}
A simple but realistic distribution for the latent or infectious periods is obtained in the limiting case of fixed periods when the survival functions are step functions, and therefore the probability density distributions are Dirac delta distributions,
$\delta(s-T_{he})$,  $\delta(s-T_{hi})$, $\delta(s-T_{ve})$. In this limiting case, the integro-differential system obtained by differentiation of the integral equations system  (see model \ref{int-diff-1}-\ref{int-diff-5} in the \ref{sec:appendixA}) reduces to a system of differential delayed equations,
}

\begin{align}
\frac{dH_s}{dt} =& \, - \beta_hV_i\frac{H_s}{H}\label{modseir-retHs} \\
\frac{dH_e}{dt} =& \,\beta_hV_i\frac{H_s}{H}- \beta_hV_i(t-T_{he})\frac{H_s(t-T_{he})}{H}\label{modseir-sei-ret}\\
\frac{dH_i}{dt}=& \,H_{i}(0)\delta(t-T)+\beta_hV_i(t-T_{he})\frac{H_s(t-T_{he})}{H} \nonumber \\
& - \beta_hV_i(t-T_{he}-T_{hi})\frac{H_s(t-T_{he}-T_{hi})}{H} \\
\frac{dH_r}{dt} =& \, \beta_hV_i(t-T_{he}-T_{hi})\frac{H_s(t-T_{he}-T_{hi})}{H}\\
\nonumber\\
\frac{dV_s}{dt}=& \, \Lambda_v-\beta_vV_s\frac{H_i}{H}-  \mu_vV_s\label{mod1eqVs}\\
\frac{dV_e}{dt}=& \, \beta_vV_s\frac{H_i}{H} -e^{-\mu_vT_{ve}}\beta_vV_s(t-T_{ve})\frac{H_i(t-T_{ve})}{H} -\mu_vV_e\\
\frac{dV_i}{dt} =& \, e^{-\mu_vT_{ve}}\beta_vV_s(t-T_{ve})\frac{H_i(t-T_{ve})}{H}
-\mu_vV_i\label{modseir-retVi}
\end{align}
where $T_{ve}$, $ T_{he}$, and $T_{hi}$ are the (fixed) latency and infectious periods of vectors and hosts. As discussed above, vector's infectious period is assumed exponentially distributed as we considered a constant vector mortality rate.
Host mortality is disregarded and then all latent host become infectious (and then $f_h = 1$). However only a fraction $f_v = e^{-\mu_v T_{ve}}$ of infected vectors survive the latency period becoming infectious.

Therefore the basic reproduction number is given by

\begin{equation}
R_0^{(3)}=\beta_h\beta_v T_{hi}\frac 1{\mu_v}e^{-\mu_vT_{ve}}\frac{V}{H}\label{R03}
\end{equation}

{\color{red}
\subsection{Relationship between the Basic reproduction numbers, and the basic modified model}

Suppose that we are studying a host-vector system for which there are estimations of the parameters as the mean latency and infectious periods. If latency periods are disregarded and we assume that all the periods are exponentially distributed, we may use the basic model (\ref{mod1Hs}-\ref{mod1Vi}) for which the basic reproduction number is

$$
R_0^{(1)}=\frac{\beta_h\beta_v}{(r_h+\mu_h)\mu_v}\frac{V}{H}.
$$

However, a more realistic model should include the latency periods. Under the most common, but unrealistic, assumption of exponentially distributed periods the corresponding model is \ref{modseirHs}-\ref{modseirVi} and the basic reproduction number is

$$ R_0^{(2)} = R_0^{(1)} \left(\frac{k_v}{k_v+\mu_v}\right)< R_0^{(1)}.$$

The delayed model (\ref{modseir-sei-ret} \ref{modseir-retVi}) is a
more realistic choice for which the basic reproduction number is

$$R_0^{(3)}=\beta_h\beta_v T_{hi}\frac 1{\mu_v}e^{-\mu_vT_{ve}}\frac{V}{H}\label{R03}.$$

Because $\frac{k_v}{k_v+\mu_v} \geq e^{-\mu_v T_{ve}}$, the basic reproduction numbers for the different models satisfy

}
\begin{equation}R_0^{(1)}>R_0^{(2)}>R_0^{(3)}.\label{desigualdad} \end{equation}

Therefore we expect larger and faster epidemics for the simple $SIR-SI$ model (Eqs. \ref{mod1Hs} - \ref{mod1Vi}) than the obtained with the more realistic models.

{\color{red}
However, it is possible to implicitly include the effect of latency in the vector population in the basic model (\ref{mod1Hs}-\ref{mod1Vi}) modifying the equation (\ref{mod1Vi}) as}

\begin{equation}
\frac{dV_i}{dt} =e^{-\mu_vT_{ve}}\beta_vV_s\frac{H_i}{H}-\mu_vV_i\label{mod1Vimod}
\end{equation}

The basic reproduction number for the {\it Basic modified model}  (Eq. \ref{mod1Vimod})  is $R_0^{(3)}$, the same as the most realistic {\it Delayed model}.

{\color{red}
\section{A stochastic agent based Ross-Macdonald model}

A stochastic version of an ordinary differential equations model
like (\ref{modseirHs}-\ref{modseirVi}) is straightforward.
Consider, for example, the simple Ross model

\begin{align}
\frac{dH_i}{dt} &=\beta_h {V_i}\frac {H_s}H -r_h H_i \nonumber \\
\frac{dV_i}{dt} &=\beta_v V_s\frac{H_i}H -\mu_v V_i \nonumber \\ \nonumber
\end{align}

In this case there are only four events: host infection, vector infection, human
recovery, and vector death. The rates of the deterministic model
define the probabilities of occurrence of the events per unit of
time or transition rates. Thus, for example, probability of human
infection in an interval $\delta t$ is given by

$$
P(H_s\rightarrow H_i,\delta t)= \beta_h {V_i}\frac {H_s}H \delta t+ o(\delta t)
$$

\noindent where $o(\delta t)$ are higher order terms for which  $\displaystyle\lim_{\delta
t\rightarrow 0 } \frac{o(\delta t)}{\delta t}=0$.

The interval between consecutive events is exponentially
distributed with parameter equal to the sum of the all transition
rates. This kind of stochastic models are markovian, the
probability of occurrence of any event depend only of the present
values of the variables.

A stochastic version of the integral Volterra  equations model
like (\ref{volterra1}-\ref{volterra5}) is not that easy (see for
example \cite{mohammadi2016}) as the corresponding stochastic model is
non-markovian, the dynamics depends of the history of the system.
One alternative is to consider  Gamma distributed periods with
integer shape parameter values, for which the equivalent systems
becomes markovian and therefore it is possible to use the
stochastic simulation scheme outlined above.

We preferred to develop an agent based model (ABM) for which the
simulation of periods with arbitrary distributions is
straightforward.} Agent based models are a computational
tool which allows to simulate populations dynamics considering the
features of each individual in the population and the interaction
between them  \cite{nepomuceno2016}. Agent based models are
considered the most realistic models where some features, like the
mobility of each individual, can be easily incorporated
\cite{bian2004, otero2011}.

{\color{red}
Our model considers a SEIR model for the host population and a SEI model for
the vector population. 

\subsection{Modeling disease transmission, progression and recovery.}

For each host and vector, the followings
attributes were considered:
\begin{itemize}
\item The epidemiological status ({\it State}) which may take  the values  {\it SUSCEPTIBLE, EXPOSED, INFECTIOUS, RECOVERED}.

\item The age of infection (the time elapsed from first infection).

\item The age of infectiousness (the time elapsed since progression to the infectious status).
\end{itemize}

} 
We considered a fixed time step $\Delta t$.

For
each vector we generated a (pseudo)random number $u$ with uniform
distribution in the interval (0,1). If $u<1-exp(-b\Delta t)$ the
vector bites a host selected at random which may be infected (if
the vector is infectious and the host susceptible) or transmit the
infection (if the vector is susceptible and the host infectious)
with probabilities $p_{h}$ and $p_{v}$ respectively.

Latent or infectious periods are random variables with some
probability distribution. We considered the general and flexible
case where periods are Gamma distributed. Special cases of the
Gamma distribution include exponential distribution (for the shape
parameter $k=1$) and the limiting case of fixed period
($k\rightarrow\infty$).

{\color{red} Waiting periods in the different epidemiological classes were
simulated in the following way. For each newly infected individual
we drawn a pseudo random number from the corresponding Gamma
distribution. This simulated value of the latency period ($\tau_e$), plus the current time $t$, was stored in the variable $T\_change$. In all cases $T\_change$ is a future time at which the individual will change the epidemiological status.

When the age of infection becomes greater or equal to $\tau_e$ we
changed the agent's state from {\it EXPOSED} to {\it INFECTIOUS} (in our implementation this is equivalent to the condition $t\geq T\_change$). In a similar
way, for each newly infectious individual we drawn a value for the
infectious period ($\tau_i$) and stored it in $T\_change$ (as before $T\_change$ is the simulated infectious period plus the actual time $t$). When the age of
infectiousness reach (or surpass) this value the state of the
individual was changed from {\it INFECTIOUS}  to {\it RECOVERED}. For the
cases of fixed waiting times the transitions are deterministic and
are determined by the values selected for the different (fixed)
periods.
}

{\color{red} Other distributions  for the latent and infectious
periods may be easily incorporated as long a generator of random
numbers for the corresponding distribution is available.}

{\color{red}
\subsection{Modeling births and deaths.}

We disregarded host births and deaths. For the vector population we considered a constant mortality
($\mu_v = 1/T_v$) and a constant birth rate $\Lambda_v$. Thus, the
probability of a vector dying in a time interval $\Delta t$ is
equal to $1 - e^{-\mu_v  \Delta_t}$. On the other hand, the number
of newborns vectors in a time step $\Delta t$ was modelled by a
Poisson random variable with parameter $\Lambda_v \Delta t$. }

\subsection{Simulation procedure}

The simulation procedure used is described in the following pseudo-code:

\begin{enumerate}

\item Initialization of variables and parameters
\begin{enumerate}
\item Set the host ($H(0)$) and vector ($V(0)$) population sizes, and the initial conditions $H_s(0)$, $H_e(0)$, $H_i(0)$, $H_r(0)$, $V_s(0)$, $V_e(0)$, $V_i(0)$.

\item Set the time step $\Delta t$, the simulation duration $t_{sim}$ and the current time $t$ equal to 0.

\item Set the values of parameters $\mu_v$, $p_v$, $p_h$, $k_v$, $k_h$, $\gamma_h$, $b$.
\end{enumerate}

\item While $t \leq t_{sim}$ and $ 0 \leq H_e(t) + H_i(t) + V_e(t) + V_i(t)$ /* \textrm{this last sentence interrupts the program when infections cannot takes place anymore } */

\begin{enumerate}

\item A random number of susceptible vector are added to the population according to a Poisson distribution with parameter $
\mu_v \Delta t$

\item For each vector in the population

\begin{enumerate}
\item A uniform random number {\color{red} in the interval $(0,1)$} is generated.

\item If the number is less than or equal to $b\Delta t$, the vector bites.
\begin{enumerate}
\item[] The host bitten is chosen at random.

\item[] If the vector is susceptible and the host bitten is infected

\begin{itemize}
\item[] A uniform random number {\color{red} in the interval $(0,1)$} is generated.

\item[] If the number is less than or equal to $p_v$, the mosquito becomes exposed

\begin{itemize}
\item[] $vector.State$ = EXPOSED, ${V_s(t) = V_s(t) -1}$, ${V_e(t) = V_e(t) + 1}$

{\color{red} \item[] Generate a latency period $\tau_e$ according to the corresponding Gamma distribution.
\item[] Set an exposed time $vector.T\_change=t+\tau_e$.}
\end{itemize}

\end{itemize}

\item[] If the vector is infected and the host bitten is susceptible

\begin{itemize}
\item[] A uniform random number {\color{red} in the interval $(0,1)$} is generated.

\item[] If the number is less than or equal to $p_h$, the host becomes exposed

\begin{itemize}
\item[] $host.State$ = EXPOSED, $H_s(t)=H_s(t)-1$, ${H_e(t)=H_e(t)+1}$

{\color{red} \item[] Generate a latency period $\tau_e$ according to the corresponding Gamma distribution.
\item[] Set an exposed time $host.T\_change=t+\tau_e$.}

\end{itemize}

\end{itemize}

\end{enumerate}

\item A uniform random number {\color{red} in the interval $(0,1)$} is generated.

\item If the number is less than or equal to $1 - e^{-\mu_v  \Delta_t}$

\quad The vector dies and it is removed from vector population.

\item Else

\begin{enumerate}

\item[] If the vector is exposed and $vector.T\_change$ is {\color{red} less than or equal to the current time} $t$

\quad $vector.State$ = INFECTED, $V_e(t)=V_e(t)-1$, ${V_i(t)=V_i(t)+1}$

\end{enumerate}

\end{enumerate}

\item For each host

\begin{enumerate}

\item If the host is exposed and $host.T\_change$ {\color{red} is less than or equal to the current time} $t$

\quad $host.State$ = INFECTED, $H_e(t)=H_e(t)-1$, ${H_i(t)=H_i(t)+1}$

{\color{red} \item[] Generate a infectious period $\tau_i$ according to the corresponding Gamma distribution.
\item[] Set an infectious time $vector.T\_change=t+\tau_i$.}

\item If the host is infected and $host.T\_change$
is {\color{red} less than or equal to the current time} $t$

\quad $host.State$ = RECOVERED, $H_i(t)=H_i(t)-1$, ${H_r(t)=H_r(t)+1}$

\end{enumerate}

\end{enumerate}

\end{enumerate}

\section{Some numerical results}

The simulations start with one host infectious, and all the other individuals susceptible. We used the day as the unit of time.

{\color{red}

\subsection{Vector to host ratio} 

Host population was
constant along the simulations. For the vector population  we
considered two cases, constant populations and seasonal varying
populations. In the first case, the population may fluctuate
stochastically about the deterministic equilibrium which was the
initial population in our simulations. In the second the
population presents seasonal oscillations around this value. The
ratio vector per host was set in 1. This value vary from system to
system and along the year in seasonal environments.  For \textit{Aedes
aegypty} for example values of 0.5 and 1.1 per human were estimated
\cite{romerovivas2005, focks2000}. Studies in malaria estimated
the number of anophelines per person in the range 3 to 4
depending on the location \cite{macdonald1955}. A typical
household in high risk areas of {\it T. cruzi} transmission has an
average of 5 human host while triatomine population may vary from
some few individuals to several hundreds \cite{nouvellet2013,
tomasini2017}.

\subsection{Parameter values}

Vectors life expectancy was set in 10 days, which is of the order of values obtained for most species of mosquito \cite{chitnis2008} and sandflies \cite{belen2006}. Kissing bugs has a longer life expectancy of about 5 months \cite{rabinovich1972}.

Probabilities of transmission ($p_{h}$, $p_{v}$) are in general
asymmetrical and there is a wide range of variation. For malaria
those probabilities were estimated in the order of 0.48 and 0.022
\cite{chitnis2008}, while for dengue are close to 1. In this work
we used $p_{hv}=p_{vh}=0.75$.

Human mean latency period is of the order of one week for dengue
and about 10 days for malaria \cite{molineaux1980}, while in the case of Chagas this period is between 4 and 15 days \cite{filigheddu2017}. In our simulations we considered a host mean latency period of 6 days.

Host mean infectious period is about 3 days for dengue \cite{otero2010}, 2 or 4
months for malaria \cite{molineaux1980}. In the case of Chagas, the acute phase is between 0 and 90 days, period in which the risk to transmit the infection is higher \cite{castanera2003}. We used a value of 5 days in our simulations.

Latent period for vectors was set in 7 days, a value about the observed in dengue infected mosquitoes.

The biting rate $b$ is a parameter harder to estimate. We
considered two values, $b=0.3/day$ and $b=0.5/day$, in order to
have two cases, low and high basic reproduction numbers. This
values are of the order observed for mosquitoes and were used for
Ross and Macdonald in their seminal works \cite{macdonald1955}.

}

\subsection{Epidemic curves}

\begin{figure}[h]
\centering
\includegraphics[width=0.5\textwidth]{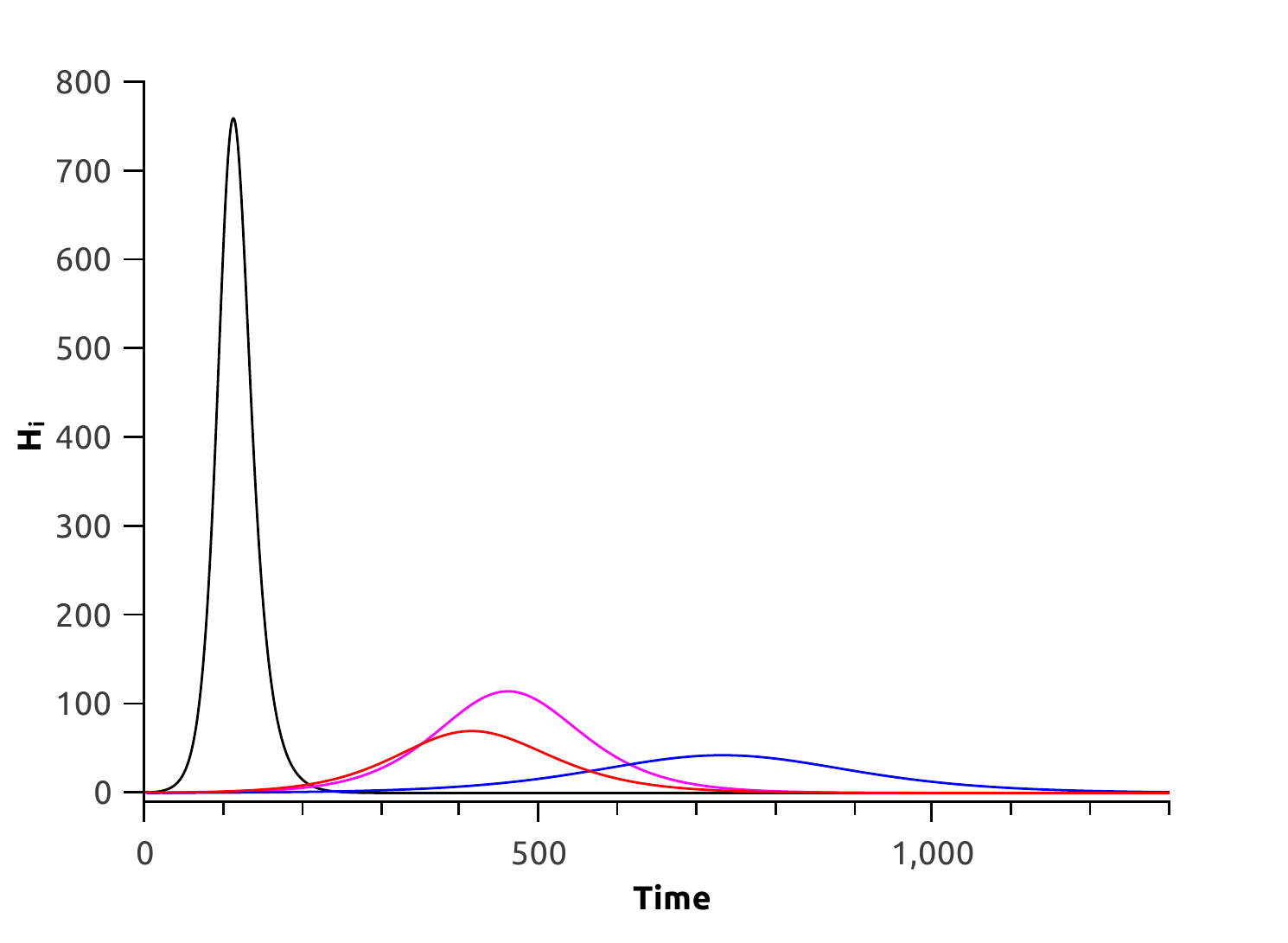}\includegraphics[width=0.5\textwidth]{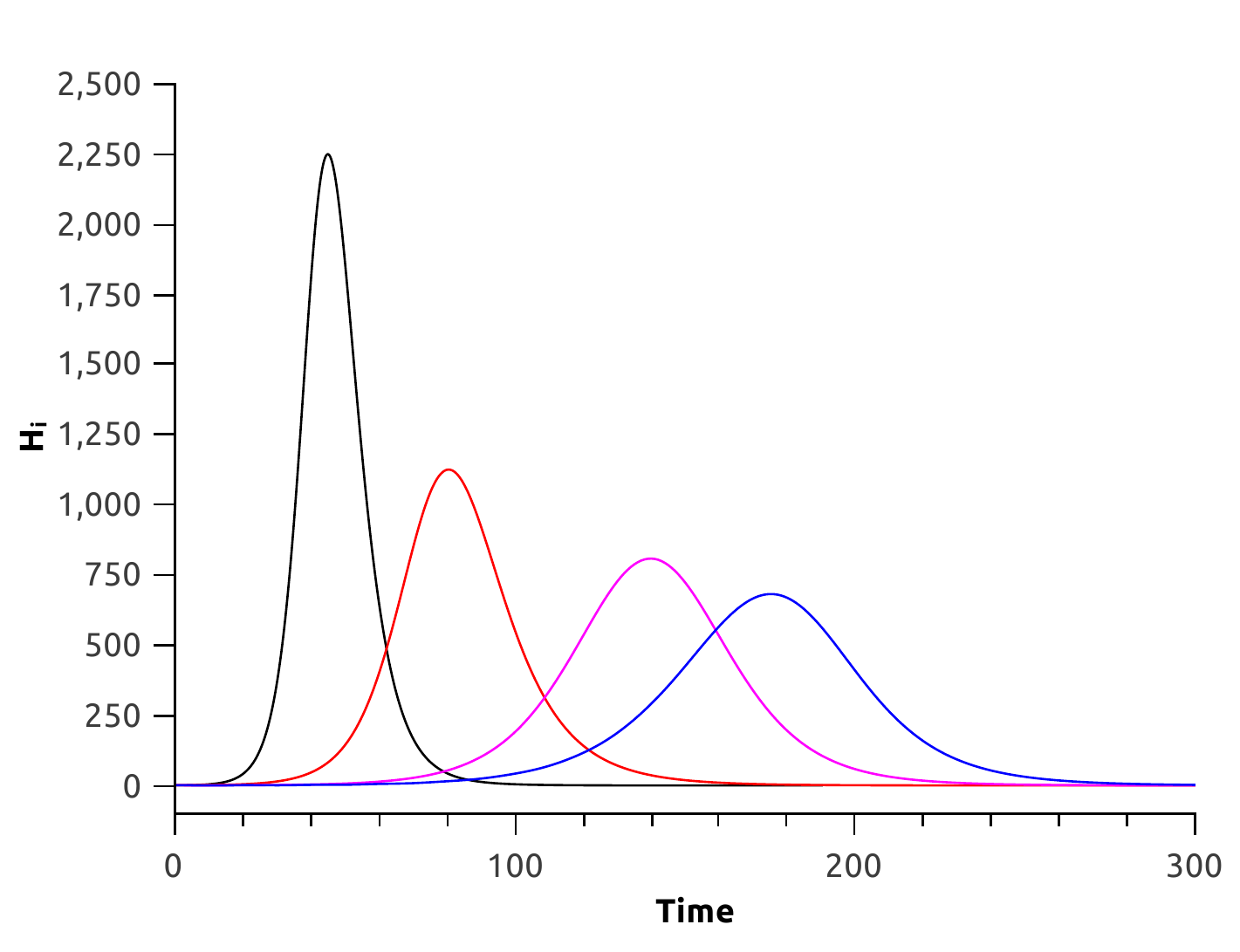}
\centering \caption{Solutions of the deterministic models (Host infectious population). Left panel: low $R_0$ ($b=0.3$), right panel, high $R_0$ ($b=0.5$). From left to right:  basic model (Eqs. \ref{mod1Hs} - \ref{mod1Vi}), basic model modified (Eq. \ref{mod1Vimod}), SEIR-SEI model (Eqs. \ref{modseirHs} - \ref{modseirVi}), delayed model (Eqs. \ref{modseir-retHs} - \ref{modseir-retVi}). Time units in days.}\label{figcompdet}
\end{figure}

In figure \ref{figcompdet} we compare numerical solutions of the models for low and high values of the basic reproduction number ($b=0.3$ and $b=0.5$, respectively) considering the parameter values in the table \ref{tab:parametros}.

\begin{table}[!h]
\begin{center}
\begin{tabular}{cc}
\hline
Parameter & Value \\ \hline
$p_v$     & 0.75  \\
$p_h$     & 0.75  \\
$T_{he}$  & 6 [days]    \\
$T_{hi}$  & 5 [days]    \\
$T_{ve}$  & 7 [days]    \\
$T_v$     & 10 [days]

\end{tabular}
\caption{Parameter values used in the simulations. In all cases we set host mortality equal to zero ($\mu_h$=0) while $\mu_v=1/T_v=$0.1 days$^{-1}$.}
\label{tab:parametros}
\end{center}
\end{table}

In table \ref{tab:CompR0} we show the corresponding $R_0$ and some statistics of the epidemic curves (number of infected host at the epidemic peak, time at which the peak is reached and the final epidemic size as proportion of the total host population size) for each of the simulations presented in figure \ref{figcompdet} corresponding to the different deterministic models.

Similar results obtained with the agent based model are presented in table \ref{tab:IBM}, where we considered only the case of exponentially distributed periods (corresponding to the SEIR-SEI model \ref{modseirHs} - \ref{modseirVi}), and fixed periods (corresponding to the Delayed model \ref{modseir-sei-ret} - \ref{modseir-retVi}).

\begin{table}[!h]
\centering
\begin{tabular}{ccccc}
\hline
Model                & $R_0$ & Epidemic Peak   & Time      & Epidemic Final Size\\
\hline
Basic model          & 2.53  & 759              & 113       & 0.86\\
Basic model modified & 1.26  & 69.7             & 415       & 0.365\\
SEIR-SEI model       & 1.49  & 114.3            & 461.25    & 0.559\\
Delayed model        & 1.26  & 42.5             & 733.4     & 0.37 \\ \hline
& & & & \\
& & & & \\
\hline
Model & $R_0$ & Epidemic Peak   & Time      & Epidemic Final Size\\
\hline
Basic model & 7.03  & 2250             & 45    & 0.99\\
Basic model modified & 3.50    & 1126          & 80        &0.925\\
SEIR-SEI model & 4.14 & 809 & 140  & 0.965\\
Delayed model & 3.50 & 682   & 175    & 0.94\\ \hline
\end{tabular}
\caption{ Basic reproduction number, peak of the epidemic,
duration from source case introduction to peak and epidemic final
size for the  solutions of the different Ross-Macdonald models
considered in figure \ref{figcompdet} left panel (top) and right
panel (bottom).} \label{tab:CompR0}
\end{table}

\begin{sidewaystable}
\centering
\begin{tabular}{cccccc}
\hline
Type of model & $b$ & \begin{tabular}[c]{@{}c@{}}Epidemic Peak\\ Mean (95\% CI)\end{tabular} & \begin{tabular}[c]{@{}c@{}}Time\\ Mean (95\% CI)\end{tabular} & \begin{tabular}[c]{@{}c@{}}Epidemic Final Size\\ Mean (95\% CI)\end{tabular} \\
\hline

\multirow{2}{*}{\begin{tabular}[c]{@{}c@{}}Exponential\\ periods\end{tabular}}
& 0.3 & 144.64 (141.36, 147.92) & 362.99 (352.1,    373.89) & 0.5622 (0.5577, 0.5666)\\
& 0.5 & 842.88 (838.60, 847.16) & 131.94 (129.80, 134.08) & 0.9664 (0.9660, 0.9668) \\
\hline

\multirow{2}{*}{\begin{tabular}[c]{@{}c@{}}Fixed\\ periods\end{tabular}}
& 0.3 & 69.48 (67.07, 71.89) & 517.91 (497.24, 538.58) & 0.3580 (0.3496, 0.3664) \\
& 0.5 & 714.5 (710.89, 718.22) & 166.30 (163.59, 169.02) & 0.9418 (0.9412,  0.9423) \\
\hline

\end{tabular}
\caption{Peak of the epidemic, duration from source case introduction to peak and epidemic final size for the  solutions of the different ABM Ross-Macdonald models considering 200 simulations.} \label{tab:IBM}
\end{sidewaystable}


Because for the same parameter values the basic reproduction number of the basic model is greater than the basic reproduction numbers of the other models (see inequality \ref{desigualdad}), the basic model produces faster epidemics with a higher epidemic final size.

For $b=0.3$ (low $R_0$'s), the epidemic final size for the basic model is about two times the epidemic final size obtained with the other models, while the epidemic peak is almost 20 times higher than the obtained with the delayed model (see table \ref{tab:CompR0}).

The modified basic model (\ref{mod1Vimod}) has the same basic reproduction number than the delayed model and both models predict almost the same final epidemic sizes. However, the first one produces higher peaks in shorter times, resulting in a epidemic that spreads through the population faster and runs out earlier. It is important to note that considering latency periods in hosts and vectors always produces lower epidemics, in comparison with the basic model.

To compare the solution of the deterministic SEIR-SEI (Eqs. \ref{modseirHs} - \ref{modseirVi}) and delayed models (Eqs. \ref{modseir-retHs} - \ref{modseir-retVi}) with the ABM results, we realized simulations following the procedure explained above considering the same parameter values (table \ref{tab:parametros}), population sizes and initial conditions used with the deterministic models.

\begin{figure}[h]
\includegraphics[width=0.5\textwidth]{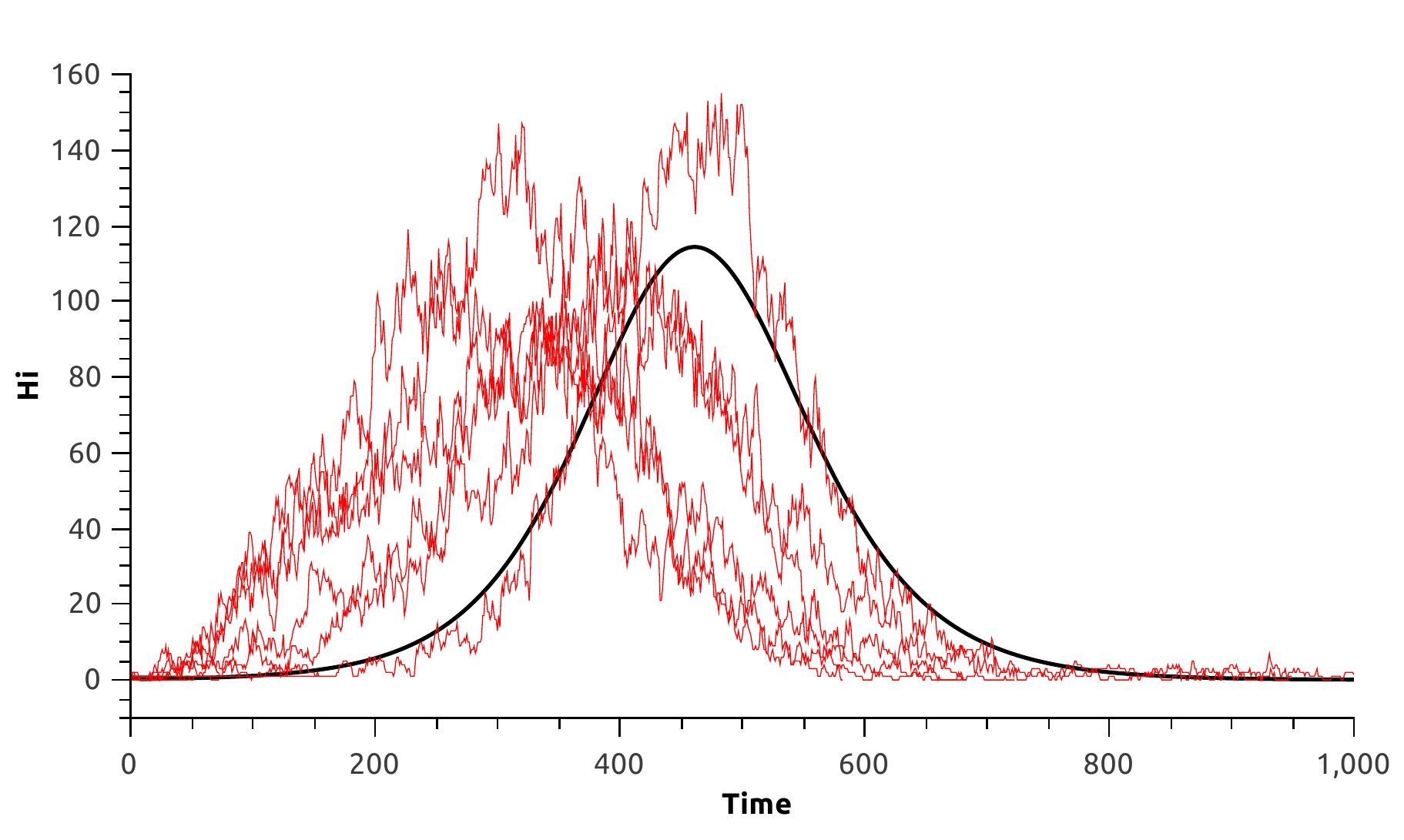}
\includegraphics[width=0.5\textwidth]{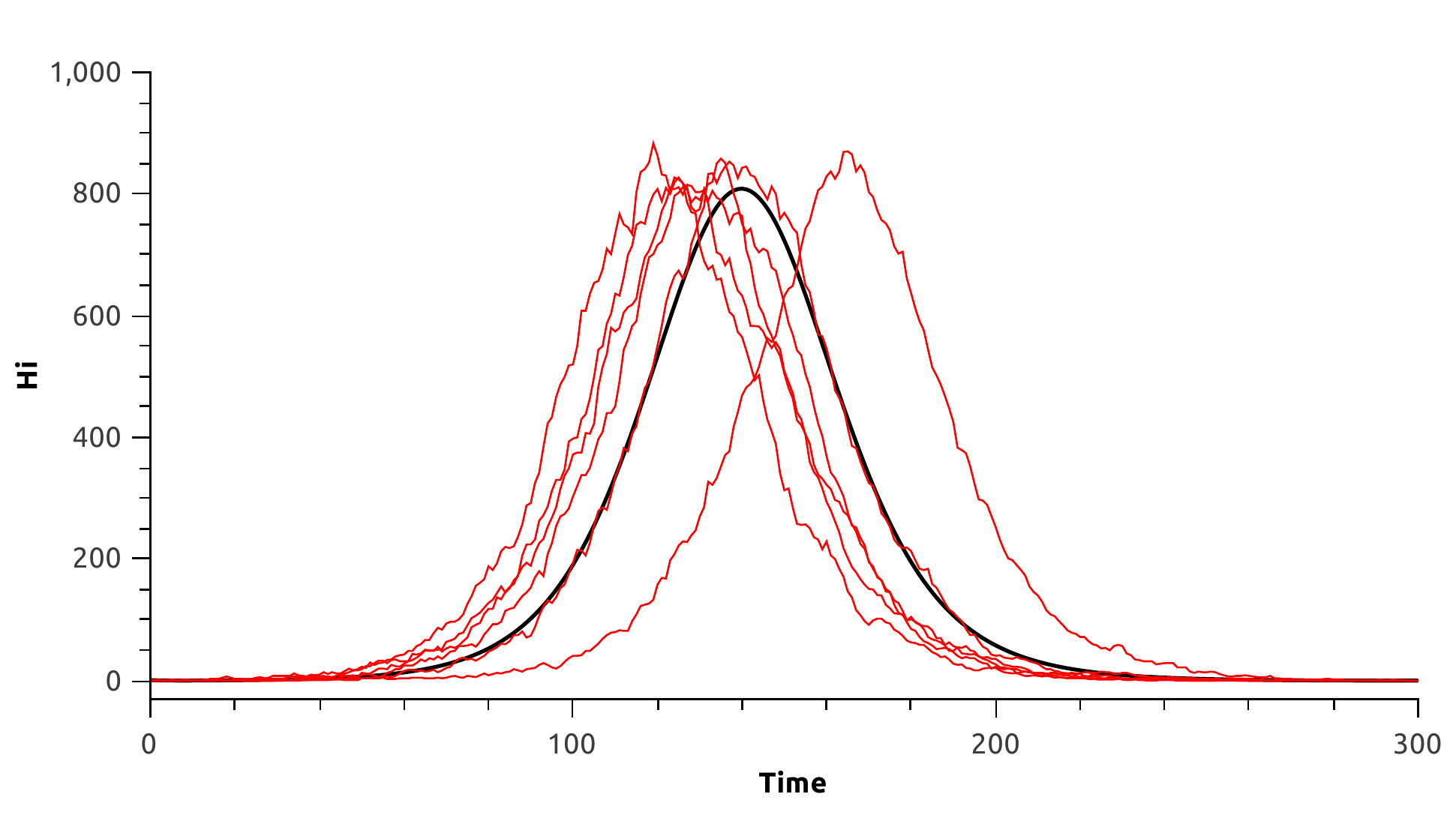}
\caption{Disease dynamics considering periods exponentially distributed and the parameters in table \ref{tab:parametros}. In black the deterministic result, and in red the ABM simulation. Left panel: low $R_0$ (b = 0.3); right panel, high $R_0$ (b = 0.5).}\label{fig:R0_exponencial}
\end{figure}

\begin{figure}[!h]
\includegraphics[width=0.5\textwidth]{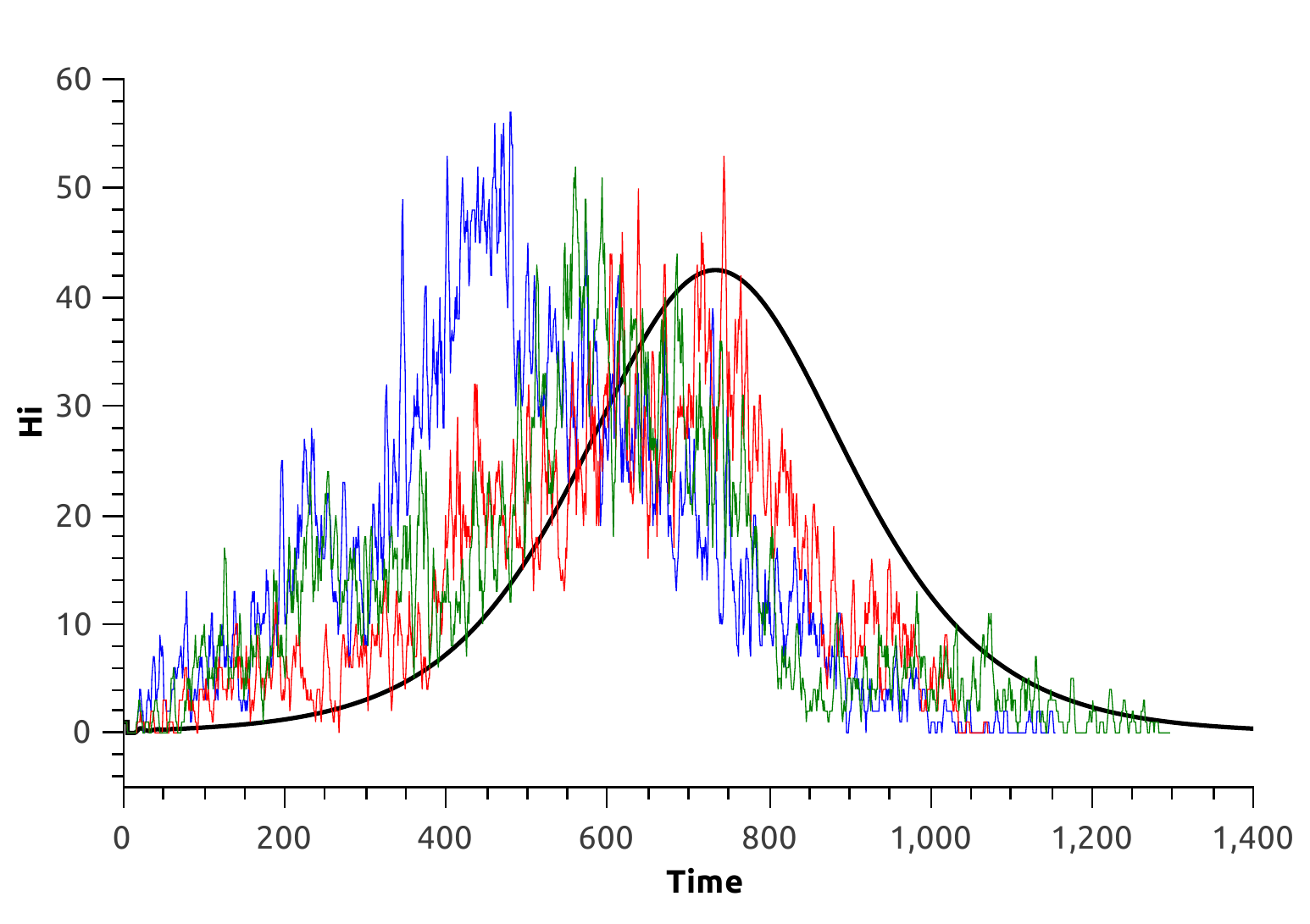}
\includegraphics[width=0.5\textwidth]{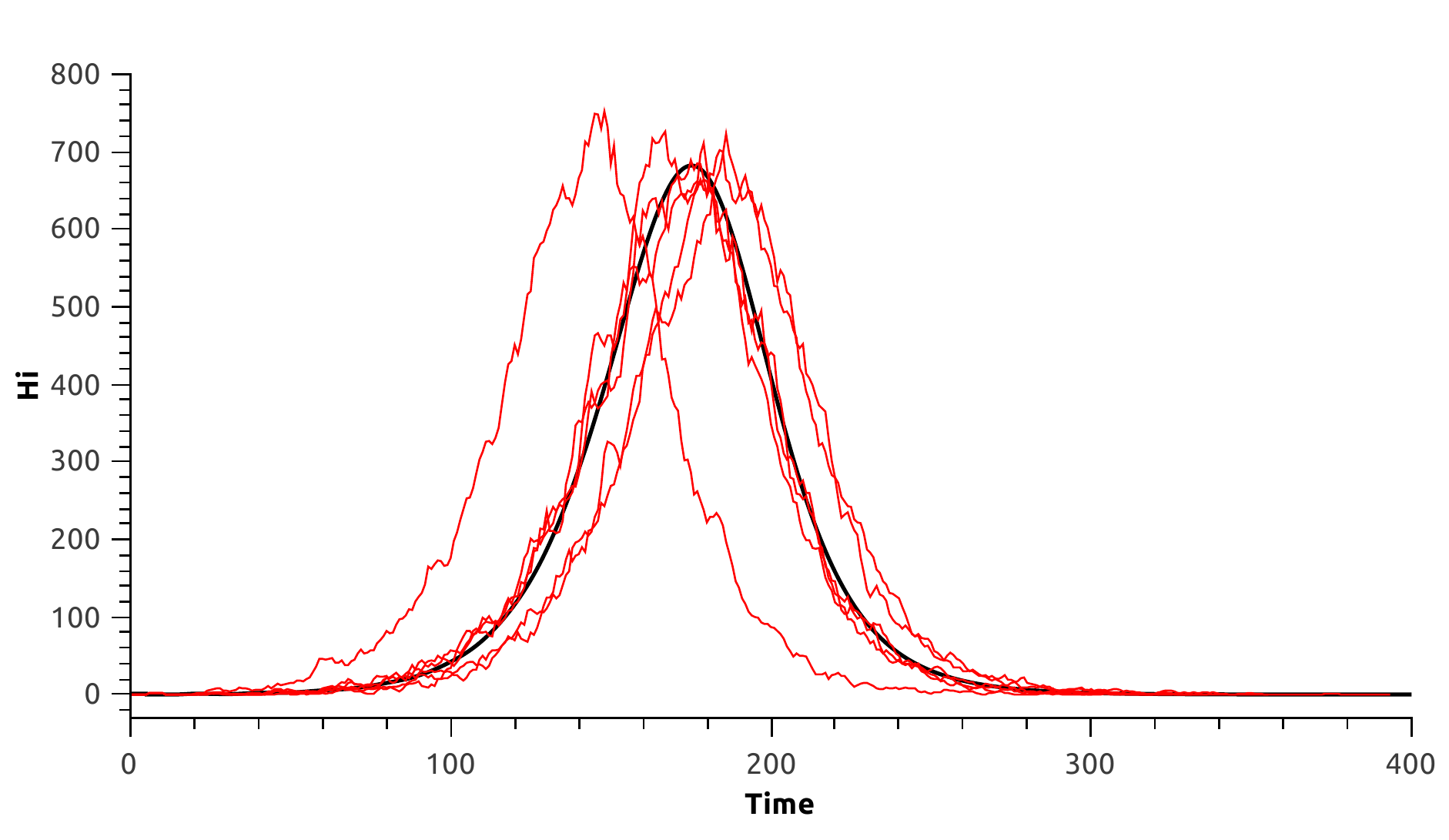}
 \caption{Disease dynamics considering fixed periods and the parameters in table \ref{tab:parametros}. In black the deterministic result, and in red the ABM simulation. Left panel: low $R_0$ (b = 0.3); right panel, high $R_0$ (b = 0.5).}\label{fig:R0_periodofijo}
\end{figure}

For $b=0.3$ (Fig. \ref{fig:R0_exponencial} - left panel and Fig. \ref{fig:R0_periodofijo} - left panel) stochasticity dominates the dynamics and realizations of the ABM are qualitatively and quantitatively different of the deterministic solutions (see table \ref{tab:IBM}). Epidemic peak is lower in the deterministic case than in the mean value obtained with the ABM simulations.

Also, the deterministic values of the epidemic peak and the time
at which it is reached are not within the corresponding 95\%
confidence intervals. Not only individual realizations are
qualitatively different from the deterministic solutions but the
mean of those realizations do not converge to the deterministic
results. 

This behaviour of the stochastic realizations is due to the fact that for low values of the basic reproduction number the stochastic effects dominate the disease dynamics. In this case stochastic dynamics is not a deterministic drift with noise \cite{aparicio2001}, and therefore we cannot expect that stochastic fluctuations average out.

For higher values of $R_0$ ($b=0.5$) stochastic dynamics is quasi
deterministic and the realizations of the ABM are similar in shape
and size to the deterministic solutions (Fig.  \ref{fig:R0_exponencial} -
right panel and \ref{fig:R0_periodofijo} - right panel). The
stochasticity may produce a shift of the epidemic curve (to the
left or to the right of the deterministic result), but it does not
greatly affect the height of the peak neither the amplitude of the
epidemic curve.

\begin{figure}[!h]
\begin{center}
\includegraphics[width=0.5\textwidth]{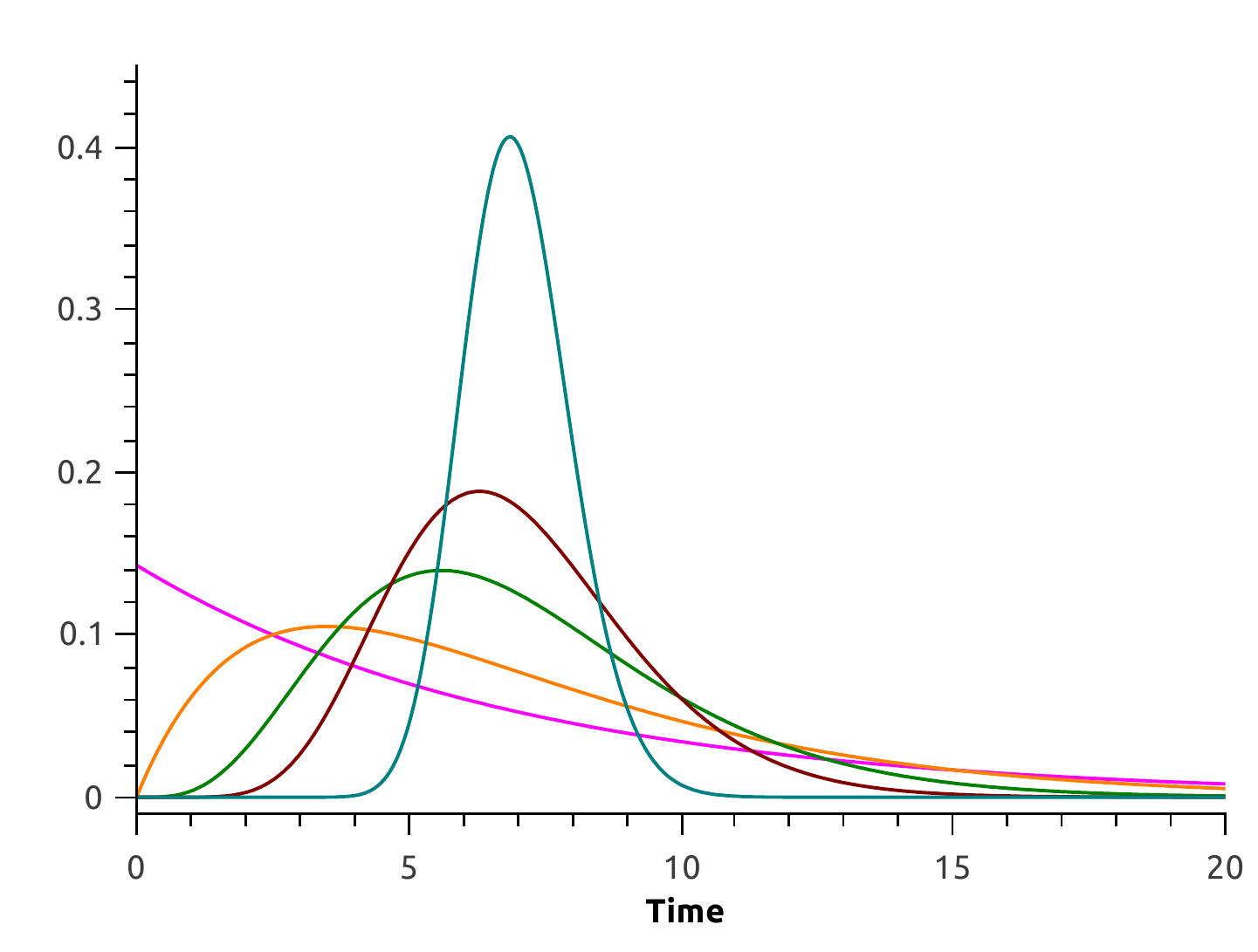}
\caption{Probability density function of Gamma distribution considering $\tau = 7$ and different values of $k$: 1, 2, 5, 10 and 50.} \label{fig:DistrosGamma}
\end{center}
\end{figure}

\begin{figure}[!h]
\includegraphics[width=0.5\textwidth]{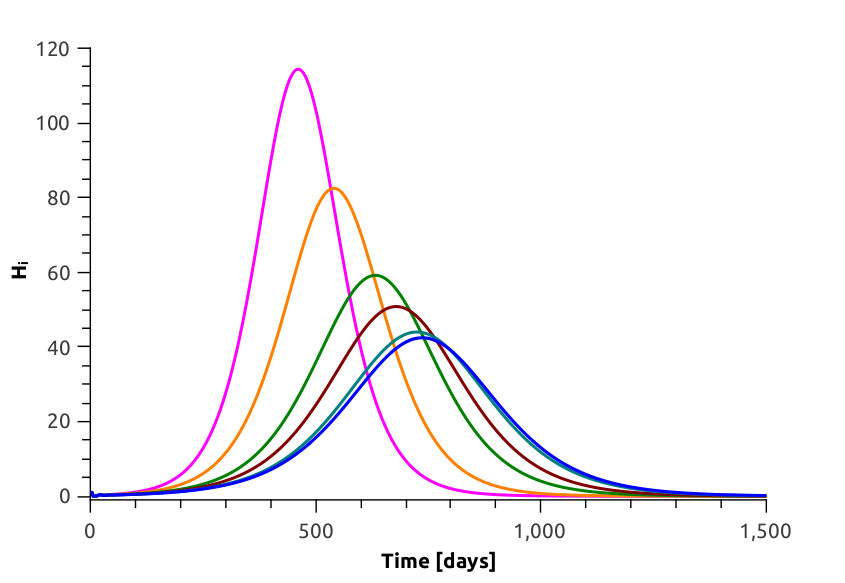}
\includegraphics[width=0.5\textwidth]{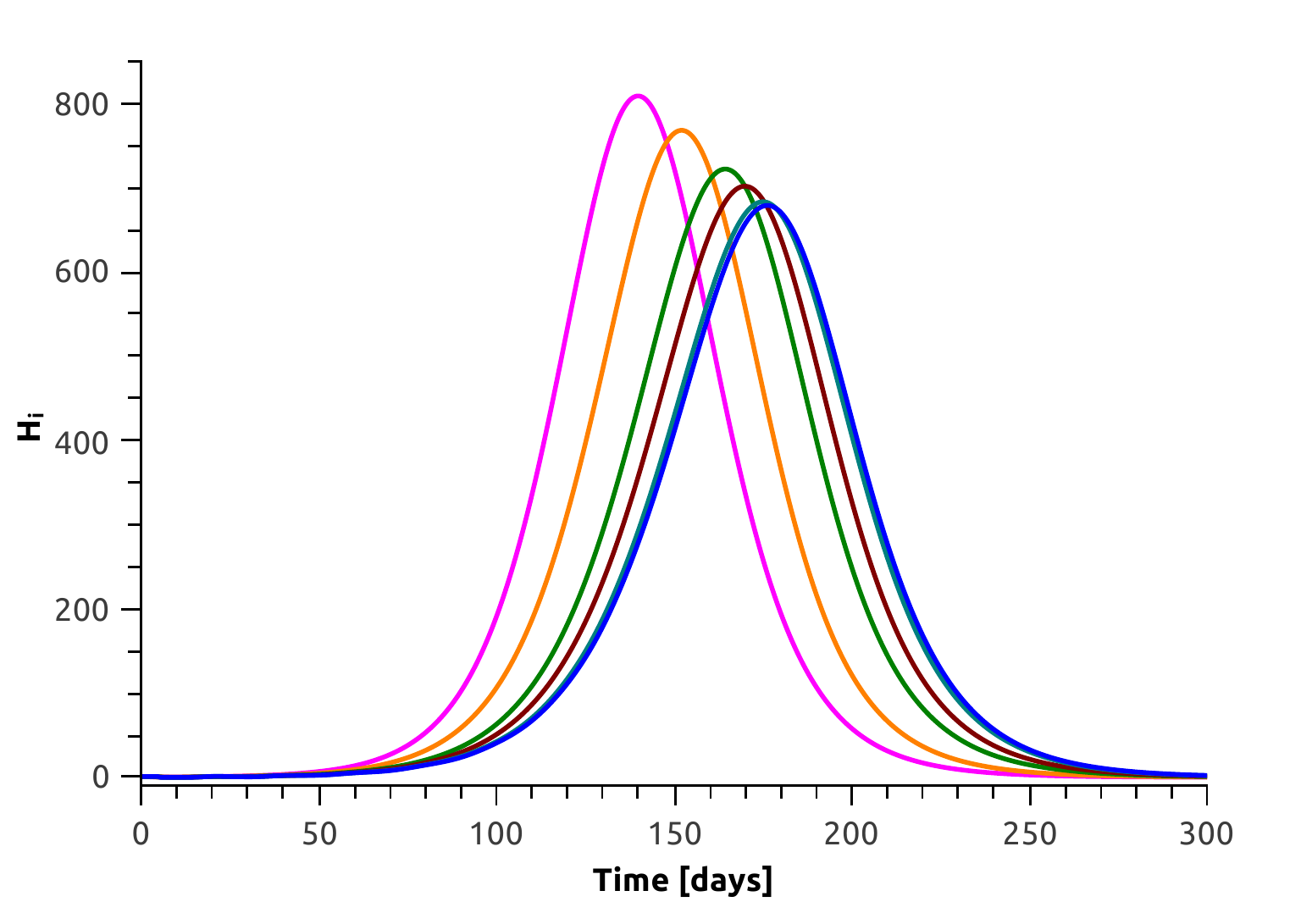}
\caption{Solutions of deterministic models (Host infectious population) considering the parameters in Table \ref{tab:parametros}. Left panel: low $R_0$ ($b=0.3$), right panel, high $R_0$ ($b=0.5$). From left to right: SEIR-SEI , model with gamma distribution for waiting periods (Eqs. \ref{Hs_gamma} - \ref{Vi_gamma}) considering $k=1$, 2, 5, 10 and 50, and delayed model (Eqs. \ref{modseir-retHs} - \ref{modseir-retVi}). For $k=1$ model (Eqs. \ref{Hs_gamma} - \ref{Vi_gamma} reduces to model (Eqs. \ref{modseirHs} - \ref{modseirVi}). 
}
\label{fig:lineartrick}
\end{figure}

{\color{red}
\subsubsection{Fixed periods vs bell shaped distributed periods}

By far the most used distribution for the waiting times is the exponential distribution, which in the deterministic case leads to model \ref{modseirHs}-\ref{modseirVi}. However, as already discussed, this is an unrealistic assumption for most cases. Latency and infectious periods are expected to have a bell shaped distribution. In the general case the model \ref{volterra1}-\ref{volterra5} should be used but numerical solutions are harder to obtain in this case. Bell shaped distributions may be modelled by the Gamma distribution, which with to independent parameters (the shape parameter $k$ and the scale parameter $\theta$) may control mean and variance (see \ref{sec:gamma_dist}).  When a Gamma distribution is used for the probability density distribution of the waiting times, a very useful property of system \ref{volterra1}-\ref{volterra5} is that for integer values of the shape parameter, the system of integral equations is equivalent to a larger system of ordinary differential equations (see \ref{sec:linear-trick}). Thus, for $k=1$, system \ref{Hs_gamma}-\ref{Vi_gamma} reduces to the SEIR-SEI model \ref{modseirHs}-\ref{modseirVi}, while for $k\rightarrow\infty$ it converges to the delayed model \ref{modseir-retHs}-\ref{modseir-retVi}.

In figure \ref{fig:DistrosGamma} we show different Gamma distributions for different values of the shape parameter $k$.  For $k=10$ there is a high variability in the waiting periods but solutions are close to the solutions obtained with fixed periods ($k=\infty$,  see fig. \ref{fig:lineartrick}). For $k=50$ both cases are almost identical.

}

\subsection{Effects of seasonality}

Seasonality is a key driver of disease dynamics in most vector-borne diseases. Seasonality affects vector population dynamics because, for example,  vector's activity is temperature dependent. In the present case we only consider that vector recruitment is affected by seasonality (mostly by variations in rainfall). As an simple example we considered harmonic variations of the form

$$
\Lambda_v=\Lambda_0[1+\epsilon sin(\omega t)].
$$

For $\epsilon =0$ we recover the case of constant recruitment used in fig. \ref{figcompdet}. For  $0< \epsilon \leq 1$ the vector population oscillates with frequency $\omega$.

\begin{figure}[!h]
\includegraphics[width=0.5\textwidth]{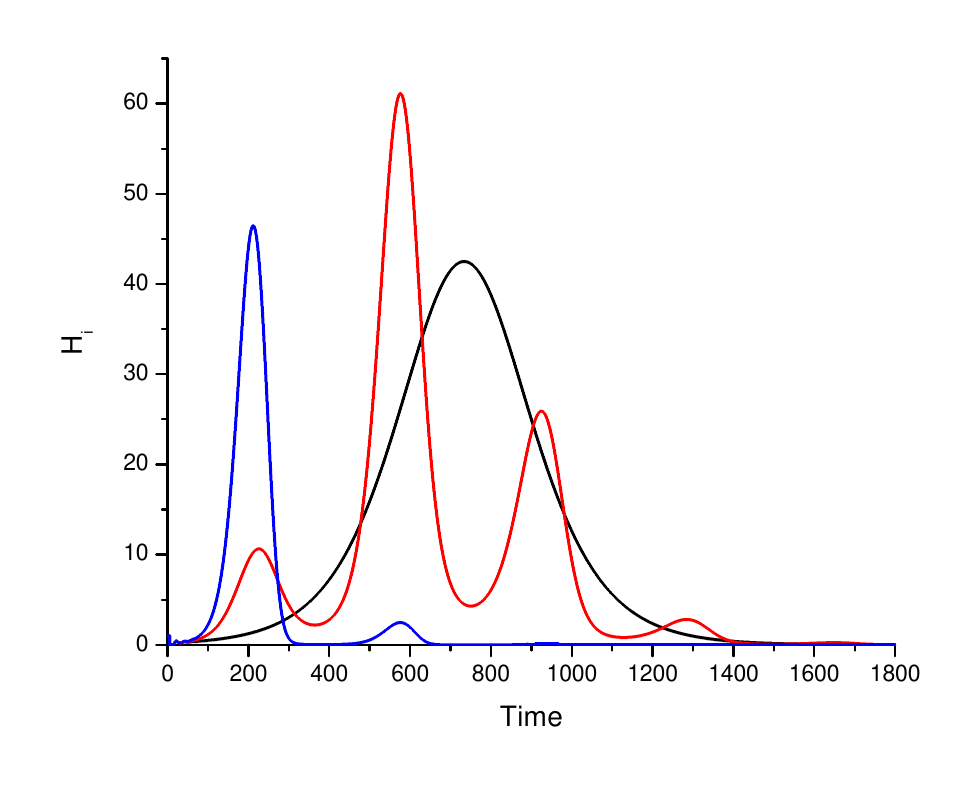}
\includegraphics[width=0.5\textwidth]{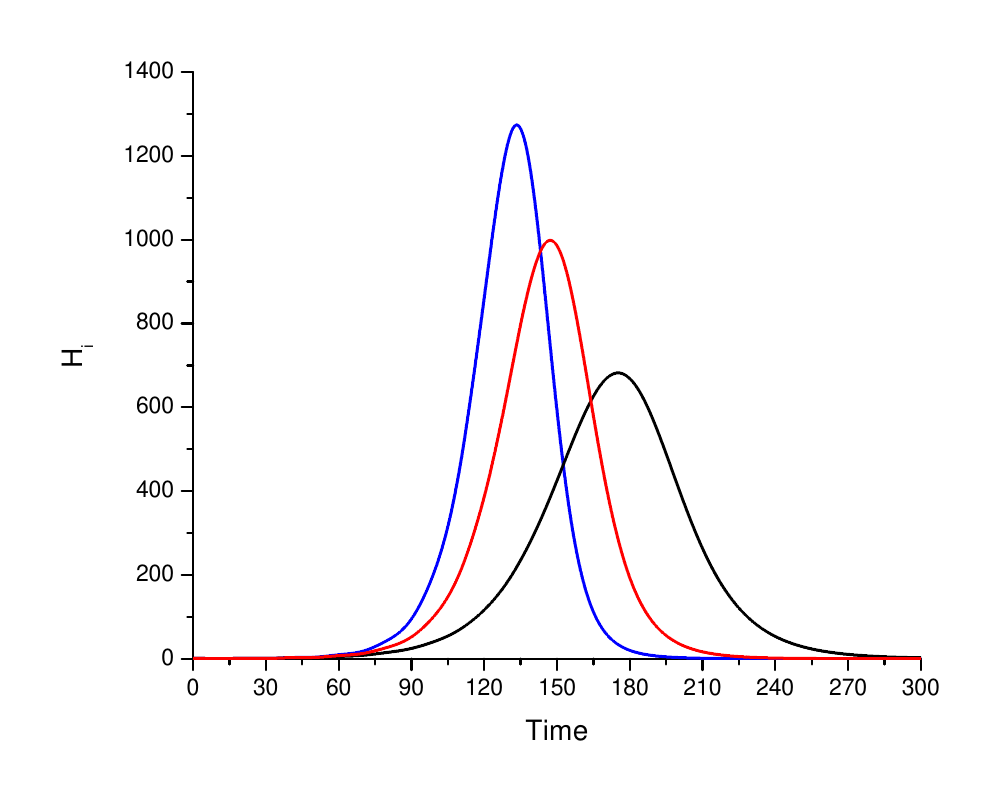}
 \caption{Numerical solutions of the delayed model  (Eqs. \ref{modseir-retHs} - \ref{modseir-retVi}) for $\epsilon=0$ (no seasonality, black line), $\epsilon =0.5$ (red) and $\epsilon = 1$ (blue), for $b=0.3$ (left panel) and $b=0.5$ (right panel).}\label{fig:season}
\end{figure}

As we show in fig. \ref{figcompdet}, duration of epidemics may last almost two years. Seasonal variation of vector populations significantly reduces epidemic duration. In fig. \ref{fig:season} we compare numerical solutions of the delayed model (Eqs. \ref{modseir-retHs} - \ref{modseir-retVi}) for different values of $\epsilon$. Duration of epidemics range between approximately 10 months to a couple of months for $\epsilon=0.5$ and $\epsilon=1$.

\begin{figure}[!h]
\includegraphics[width=0.5\textwidth]{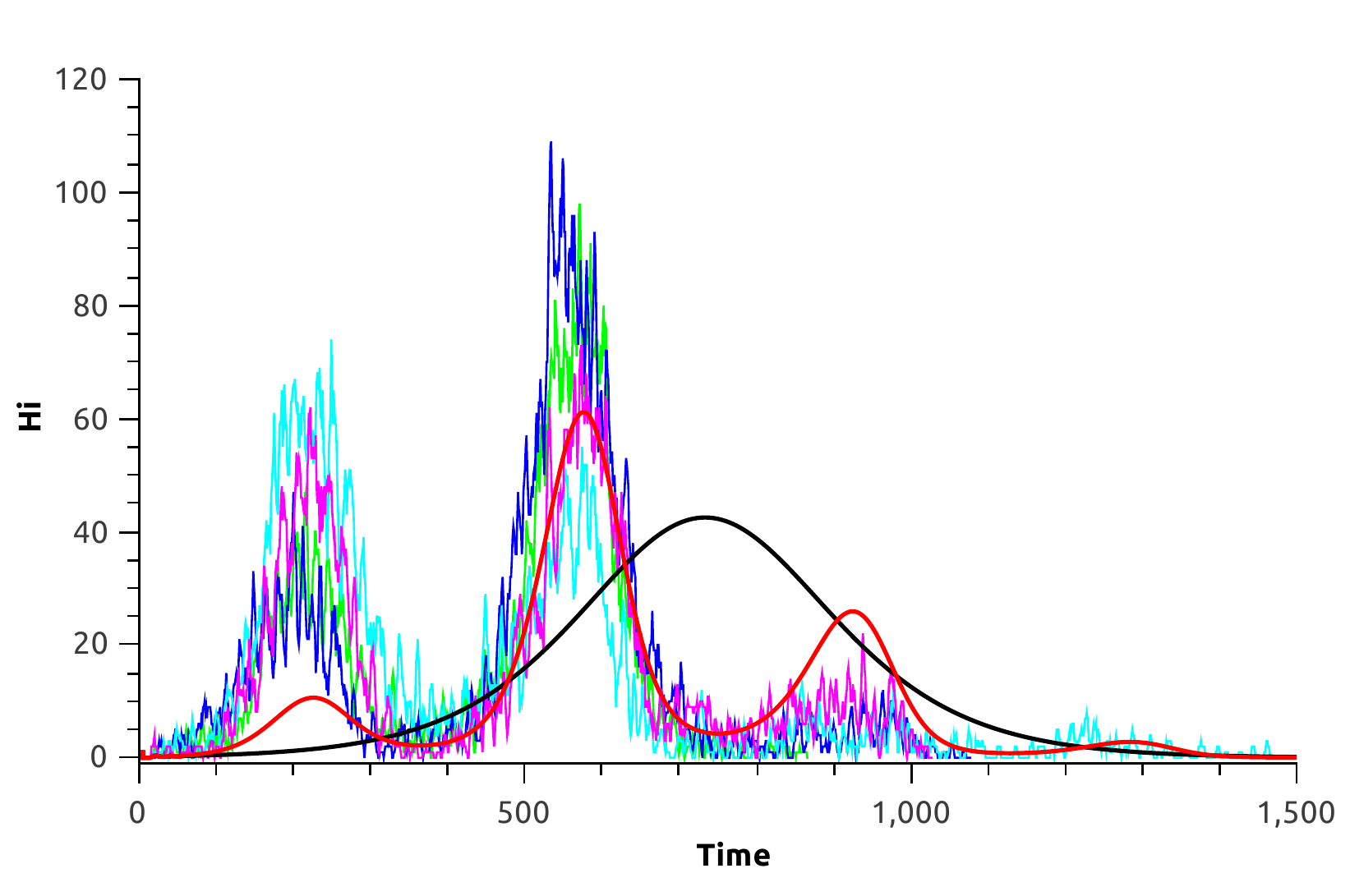}
\includegraphics[width=0.5\textwidth]{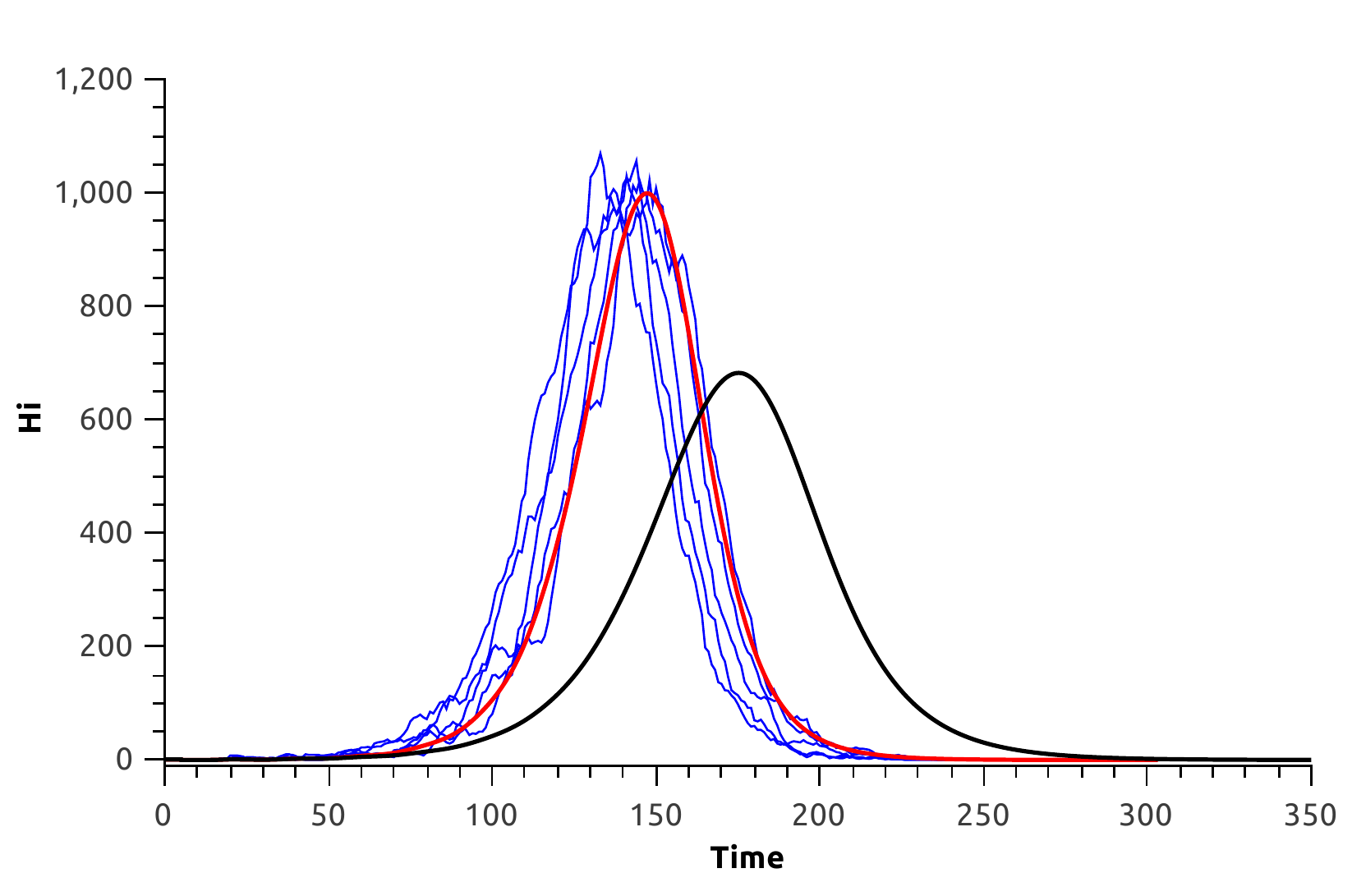}
 \caption{\color{red}{Numerical solutions of the delayed model (Eqs. \ref{modseir-retHs} - \ref{modseir-retVi}) with $\epsilon =0.5$ (red) and the corresponding ABM considering fixed periods (other colours curves), for $b=0.3$ (left panel) and $b=0.5$ (right panel). In black line the solution of the delayed model (Eqs. \ref{modseir-retHs} - \ref{modseir-retVi}) without seasonality ($\epsilon=0$).}}\label{fig:IBM-season}
\end{figure}

Fig. \ref{fig:IBM-season} shows numerical solutions of the ABM considering fixed periods and seasonality. We can see that in the case in which $R_0$ is higher ($b=0.5$) the ABM results are similar to the deterministic model. Conversely, considering $b=0.3$, the curves obtained from the AMB are qualitatively similar to the deterministic case, since the same amount of peaks can be clearly observed in both cases. However, the values reached in these peaks in the case of ABM are higher, in general, in the first and second one and lower in the third one.

\subsection{Computing $R_0$ from the agent based model}

To compute $R_0$ in the case of the agent based model, we have to follow the infectious generation of hosts and vectors. So, the procedure realized is as follow. The first infected host is the only host of first infected generation. The vectors infected by a host of first generation, are vectors of first infected generation. When a vector of first infected generation, infects a susceptible hosts, these host are second infected generation. In general, when a host of infected generation $m$ infects a vector, the infected generation of the vector is $m$. Then, when a vector of infected generation $m$ infects a host, then infected generation of the host is $m+1$.

Let $H_m$ be the number of infected-host generation $m$. Then, $R_0$ can be estimated as $R_0 \approx  H_3 / H_2$ \cite{aparicio2007}. Due to the stochasticity of the ABM simulations, it is important to realize a considerable number of simulations and then calculate the mean of the $R_0$ value estimated for each simulation. An $R_0$ estimation considering 200 realizations of the simulations analyzed en the previous section is presented in the table \ref{tab:estimacionR0}. In all the cases the deterministic value of $R_0$ is within the 95\% confidence interval.

\begin{table}[!h]
\centering
\begin{tabular}{ccccc}
\hline                                                                                              Type of model & $b$ & \begin{tabular}[c]{c}Deterministic\\ $R_0$\end{tabular} & \begin{tabular}[c]{@{}c@{}}Estimation $R_0 \approx H_3 / H_2$\\ Mean (95\% CI)\end{tabular} \\ \hline

\multirow{2}{*}{\begin{tabular}[c]{c}Exponential\\ periods\end{tabular}}
& 0.3 & 1.25 & 1.28 (1.01, 1.54)\\
& 0.5 & 3.49 & 3.60 (3.16, 4.05)\\ \hline

\multirow{2}{*}{\begin{tabular}[c]{c}Fixed \\ periods\end{tabular}}
& 0.3 & 1.49 & 1.45 (1.21, 1.69) \\
& 0.5 & 4.14 & 4.03 (3.60, 4.46) \\ \hline

\end{tabular}
\caption{Estimation of $R_0$ from the ABM model considering 200 simulations.}\label{tab:estimacionR0}
\end{table}

As we can see in Eqs. \ref{R02} and \ref{R03}, given the parameters of the host and vector populations, $R_0$ is a linear function of the relation $V/H$. So, varying the relation $V/H$, we can obtain different values of $R_0$.

\begin{figure}[h]
\centering
\includegraphics[width=0.4\textwidth]{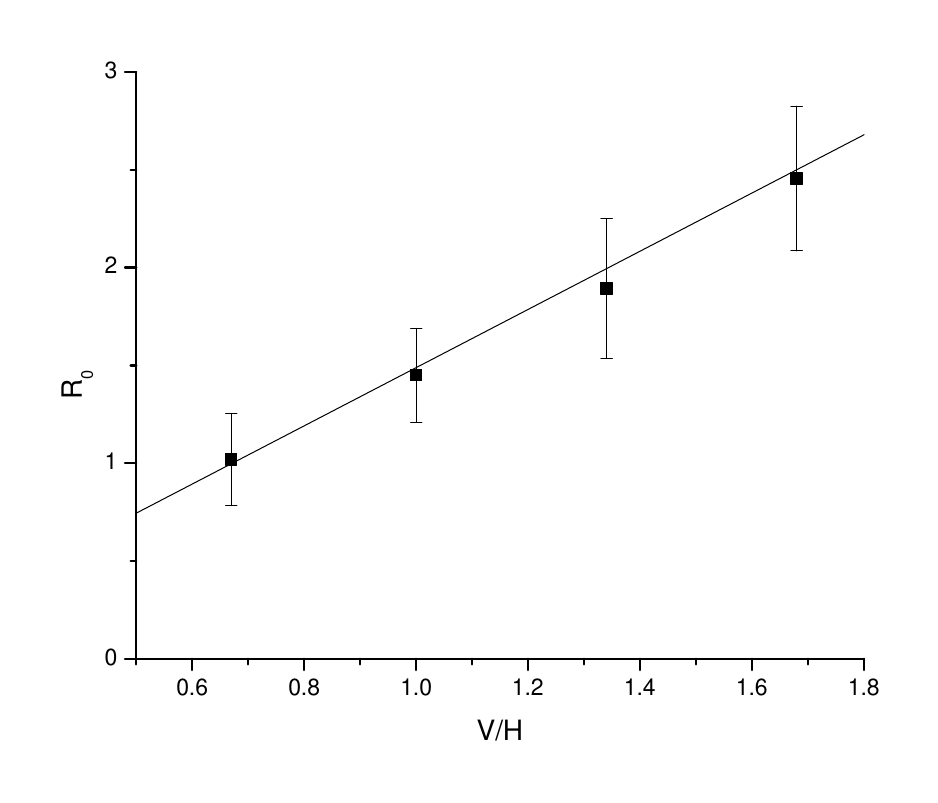}\includegraphics[width=0.4\textwidth]{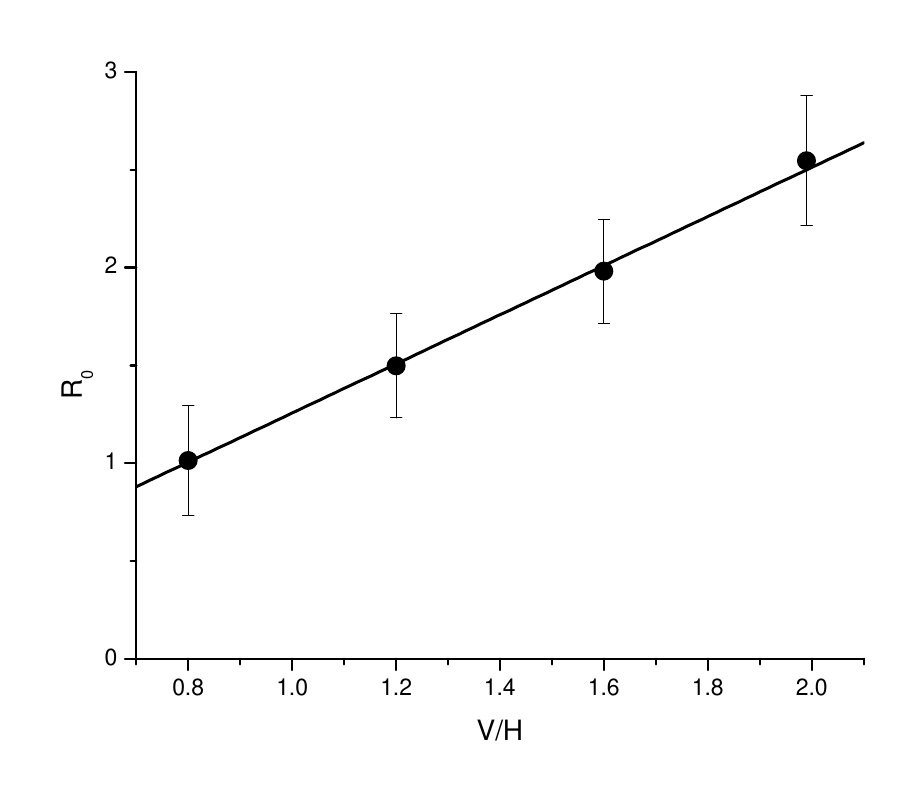}
\centering \caption{Empirical estimates of the Basic reproduction numbers (squares, bars are 95\% confidence interval) obtained with the ABM for the cases of exponentially distributed periods (left) and fixed period (right). Continuous line are the corresponding theoretical values given by expressions (\ref{R02}) and (\ref{R03}).  }\label{figR0}
\end{figure}

Considering the parameters in the table \ref{tab:parametros} and a biting rate equal to 0.3 ($b=0.3$), we estimated the value of $R_0$ from the agent based model for different values of $V/H$. The results considering exponentially distributed periods and fixed period are shown in the Fig. \ref{figR0}, respectively. In all the cases, an initial population of 10000 hosts was considered, with only one infected host. Each estimation of the basic reproduction number was realized with 200 simulations.

As can be seen in the figure \ref{figR0} numerical estimations of the basic reproductive number  are in the 95\% confidence interval estimated from the 200 ABM simulations.

\section{Discussion and Conclusions}

{\color{red} The use of mathematical models in epidemiology has a long and fruitful tradition. However different hypotheses about the systems under study may lead to, in some cases, significant different results (see for example \cite{amaku2015, wonham2006, wonham2008})}.

{\color{red}
In this work we present different formulations of the
Ross-Macdonald model using ordinary differential equations, Volterra integral equations and agent based modelling. In the
most general case we included latency periods in both vectors and
hosts. We also considered general distributions for latency and
infectious periods including two simple cases: exponentially distributed periods and fixed
periods.

As we show in this work, disregarding latency periods has a
dramatic effect in the dynamics. This is quite apparent for low
basic reproduction numbers (see Fig. \ref{figcompdet}). As in most
vector-borne diseases vector's latency periods and life expectancy
are of the same order of magnitude, disregarding latency
overestimate the basic reproduction number, and therefore we
observed faster epidemics with significantly higher peaks. A
substantial improvement is achieved with the simple modification
(\ref{mod1Vimod}) which produces the same values of $R_0$ as the
delayed model but still the epidemic curves are significantly
different.

Not only the inclusion of latency periods is important but also
its distributions. Using exponentially distributed periods leads
to slightly smaller basic reproduction numbers and still a
noticeable differences in the epidemic curves.

For the most realistic case of bell-shaped distributed waiting periods we show that numerical solutions of the Volterra integral system \ref{volterra1}-\ref{volterra5} are close to the solutions of the simpler {\it delayed model} \ref{modseir-retHs}-\ref{modseir-retVi}, for which numerical simulations may be easily obtained using a Runge-Kutta scheme, for example. For us, the delayed model is therefore the model of choice as it combines realism and simplicity.
}

A central assumption of the Ross-Macdonald models is homogeneous
random mixing: probability of biting in a susceptible host is
proportional to the fraction of susceptible host in the entire
population. This hypothesis may hold for some local, relatively
small, populations. Larger populations may be modelled using a
meta-population approach, for example. If local populations have
some degree of synchronization, the total population disease
dynamics could be quasi-deterministic (see for example
\cite{gutierrez2015}), and perhaps a Ross-Macdonald model may
describe the global dynamics of the system. In this work we
considered  populations of 10$^4$ individuals, a large enough
population for which it is not obvious that the assumption of
homogeneous mixing holds.

{\color{red} For both, high and low
values of the basic reproduction number, solutions of the
deterministic models and the estimation of the $R_0$ of the agent
based model are statistical similar (see fig. \ref{figR0} and
table \ref{tab:estimacionR0}), although the epidemic curve  may be
significantly different from the deterministic solution,
especially for low $R_0$ value (see fig. \ref{fig:R0_exponencial}
and \ref{fig:R0_periodofijo}). }

Deterministic models, like the SEIR-SEI model
\ref{modseirHs}-\ref{modseirVi}, are simple ordinary differential
equations systems with constant parameters, more amenable for
analysis. Numerical integration is straightforward using
Runge-Kutta of fourth order, for example. The more realistic
choice of fixed periods is modelled by delayed differential
equations. Analysis is more complex for these type of models but
numerical integration is easily implemented too.

For the agent based model there are not differences, neither in
the difficulty of the coding or in the computational cost for both
cases, and therefore non-exponentially distributed periods (like
fixed periods) is the recommended choice.

In our simulations we considered parameter values compatible with
some vector-borne diseases in humans like dengue. In all cases the
number of vectors per host was set equal to one at demographic
equilibrium. For low values of the basic reproduction number
epidemics obtained with the (most realistic) fixed period models
have a duration of more than two years (see Fig.
\ref{fig:R0_periodofijo}, left panel), which is never observed in
real epidemics. {\color{red} This results highlights the importance of
including seasonality when modelling some vector-borne diseases.
Vector populations usually have seasonal fluctuations, driven by
rainfall, for example, which shape the duration of the epidemics
(see fig. \ref{fig:season}). {\color{red} However, outbreaks sizes are generally a function not only of the abundance of vectors, but also of other variables such as climate, community immunity, host mobility, among others. Vector activity is strongly affected by temperature and therefore not only seasonal variations are significant but also habitat conditions as the use of air conditioning.} The two values of $\epsilon$ used correspond to moderate seasonal variations in vector abundance ($\epsilon=1/2$) as expected in endemic settings, and to marked variations in vector abundances as observed in non-endemic populations.}

{\color{red} As the homogeneous mixing assumption is expected to hold only for relatively small populations, stochasticity should be considered when modelling such cases. Larger populations may be modelled using a metapopulation approach, something we will explore in forthcoming works.

}

\section*{Acknowledgements}
This work was partially supported by grants CIUNSA 2018-2467 and
PICT 2014-2476. JPA is a member of the CONICET.
MIS is a postdoctoral fellow of CONICET.


\newpage

\appendix

\section{General SEIR-SEI model with arbitrary distributions for the waiting time} \label{sec:appendixA}

We will consider the general case of a SEIR-SEI model. For the host and vector populations we assume that the latency period ($T_{he}$ for hosts and $T_{ve}$ for vectors) and the infectious period for hosts ($T_{hi}$) are random variables with probability density distributions $f_{he}(s)$, $f_{ve}$ and $f_{hi}(s)$, respectively. The cumulative distributions are denoted by $F_{he}(s)$, $F_{ve}$ and $F_{hi}(s)$, respectively. The complementary cumulative distribution, $\bar{F_*}(s)=1-F_*(s)$, is known as the survival function and gives the probability that an individual infected in $t=0$ remains infected (or exposed, depending of the case) at time $s$.

Therefore the evolution of the populations can be described by the integral Volterra equations  \cite{feng2007}:

\begin{align}
H_s(t) &=H_s(0)-\int_0^t \beta_h V_i(s)\frac{H_s(s)}{H}ds  \\
H_e(t) &= H_e(0)\bar{F}_{he}(t)+\int_0^t \beta_h V_i(s)\frac{H_s(s)}{H} \bar{F}_{he}(t-s)ds   \\
H_i(t) &= H_i(0) \bar{F}_{hi}(t) + \int_0^t \int_0^\tau \beta_h  V_i(s)\frac{H_s(s)}{H}\left[-\frac{d\bar{F}_{he}}{dt}(\tau -s)\right]\bar{F}_{hi}(t-\tau) \; ds d\tau \\
H_r(t) &= H-H_s(t)-H_i(t)-H_e(t) \\ \nonumber
\\
V_s(t) &= V_s(0) e^{-\mu t} + \int_0^t \Lambda_v e^{-\mu(t-s)} ds - \int_0^t \beta_v V_s(s)\frac{H_i(s)}{H}ds \\
V_e(t) &= V_e(0) e^{-\mu t} + \int_0^t \beta_v V_s(s)\frac{H_i(s)}{H} \bar{F}_{ve}(t-s) e^{-\mu(t-s)}ds \\
V_i(t) & = V_i(0) e^{-\mu t} + \int_0^t \int_0^\tau \beta_v V_s(s)\frac{H_i(s)}{H} \left[-\frac{d\bar{F}_{ve}}{dt}(\tau -s)\right] e^{-\mu(t-s)} \; ds d\tau.
\end{align}

\noindent Differentiation of these equations leads to the following system of integro-differential equations,

\begin{align}
\frac{dH_s}{dt} = & - \beta_h V_i(t) \frac{H_s(t)}{H} \label{int-diff-1} \\
\frac{dH_e}{dt} = & -H_e(0)f_{he}(t) + \beta_h V_i(t)\frac{H_s(t)}{H}-\int_0^t \beta_h V_i(s)\frac{H_s(s)}{H} f_{he}(t-s)ds \\
\frac{dH_i}{dt} = & -H_i(0)f_{hi}(t) + \int_0^t\beta_h V_i(s)\frac{H_s(s)}{H}f_{he}(t-s)ds \nonumber \\
& - \int_0^t \left( \int_0^\tau \beta_h V_i(s)\frac{H_s(s)}{H} f_{he}(\tau - s) ds  \right) f_{hi}(t-\tau) d\tau\\
\frac{dH_r}{dt} = & \; H_e(0)f_{he}(t) +H_i(0)f_{hi}(t) \nonumber \\
& + \int_0^t \left( \int_0^\tau \beta_h V_i(s)\frac{H_s(s)}{H} f_{he}(\tau - s) ds  \right) f_{hi}(t-\tau) d\tau \\ \nonumber
\\
\frac{dV_s}{dt} = & \; \Lambda_v -   \beta_v V_s(t)\frac{H_i(t)}{H} - \mu V_s(t) \\
\frac{dV_e}{dt} = & \; - \mu e^{- \mu t} V_e(0) + \beta_v V_s(t)\frac{H_i(t)}{H} \nonumber \\
& \; - \int_0^t \beta_v V_s(s)\frac{H_i(s)}{H} f_{ve}(t-s) e^{-\mu(t-s)} ds  - \mu V_s(t) \\
\frac{dV_i}{dt} = & - \mu e^{- \mu t} V_i(0) + \int_0^t \beta_v V_s(s)\frac{H_i(s)}{H} f_{ve}(t-s) e^{-\mu(t-s)} ds - \mu_v V_i(t) \label{int-diff-5}
\end{align}

\section{Model with Gamma distributions for the waiting periods} \label{sec:linear-trick}

If the waiting periods are Gamma distributed with mean $\tau$ and integer shape parameter $k$, we can apply the {\it linear trick } \cite{smith2011} to solve the system of integro-differential differential equations (\ref{int-diff-1}-\ref{int-diff-5}).

Suppose that the latency and infectious periods for hosts  and the latency period for vectors are all Gamma distributed with means $\tau_i$ and integer shape parameters $k_i$ ($i=1,2,3$ respectively). Then it is possible to divide the exposed human population  in $k_{1}$ compartments such that $H_e = \displaystyle \sum_{j=1}^{k_{1}} H_{e,j}$. In a similar   fashion, the $H_i$ and $V_e$ populations may be divided in $k_2$ and $k_3$ clases respectively.

Then, the system of integro-differential equations is equivalent to the following system of ordinary differential  equations,

\begin{align}
\frac{dH_s}{dt} &=\Lambda_h-\beta_hV_i\frac{H_s}{H}-\mu_hH_s \label{Hs_gamma} \\ \nonumber  \\
\frac{dH_{e,1}}{dt} &=\beta_hV_i\frac{H_s}{H}- \frac{k_1}{\tau_1} H_{e,1} - \mu_h H_{e,1} \\
\frac{dH_{e,2}}{dt} &= \frac{k_1}{\tau_1}(H_{e,1} - H_{e,2}) - \mu_h H_{e,2}\\
 & \vdots \nonumber \\
\frac{dH_{e,k_1}}{dt} &= \frac{k_1}{\tau_1}(H_{e,k_{1}-1} - H_{e,k_{1}}) - \mu_h H_{e,k_{1}}\\  \nonumber \\
\frac{dH_{i,1}}{dt}&= \frac{k_{1}}{\tau_{1}}H_{e,k_{1}} - \frac{k_{2}}{\tau_{2}}H_{i,1} - \mu_h H_{i,1} \\
\frac{dH_{i,2}}{dt} &= \frac{k_{2}}{\tau_{2}}(H_{i,1} - H_{i,2}) - \mu_h H_{i,2}\\
 & \vdots \nonumber \\
\frac{dH_{i,k_{2}}}{dt} &= \frac{k_{2}}{\tau_{2}}(H_{i,k_{2}-1} - H_{i,k_{2}}) - \mu_h H_{i,k_{2}}\\ \nonumber \\
\frac{dH_r}{dt}&= \frac{k_{2}}{\tau_{2}}H_{i,k_{2}} -\mu_hH_r\\
 \nonumber \\
\frac{dV_s}{dt}&=\Lambda_v-\beta_vV_s\frac{H_i}{H}-\mu_vV_s\\ \nonumber \\
\frac{dV_{e,1}}{dt} &=\beta_hV_s\frac{H_i}{H}- \frac{k_{3}}{\tau_{3}} V_{e,1} - \mu_v V_{e,1} \\
\frac{dV_{e,2}}{dt} &= \frac{k_{3}}{\tau_{3}}(V_{e,1} - V_{e,2}) - \mu_v V_{e,2}\\
 & \vdots \nonumber \\
\frac{dV_{e,k_{3}}}{dt} &= \frac{k_{3}}{\tau_{3}}(V_{e,k_{3}-1} - V_{e,k_{3}}) - \mu_v V_{e,k_{3}}\\ \nonumber \\
\frac{dV_i}{dt}&= \frac{k_{3}}{\tau_{3}}V_{e,k_{3}} - \mu_v V_i \label{Vi_gamma}
\end{align}

\noindent where $H_e = \displaystyle \sum_{j=1}^{k_{1}} H_{e,j}$; $H_i = \displaystyle \sum_{j=1}^{k_{2}} H_{i,j}$, and $V_e = \displaystyle \sum_{j=1}^{k_{3}} V_{e,j}$

The Gamma distribution with shape parameter $k=1$ is an exponential distribution, while for $k \rightarrow \infty$, the Gamma distribution tends to a Dirac delta distribution. Thus, as $k$ increases from 1 to $\infty$, the model  (Eqs. \ref{Hs_gamma} - \ref{Vi_gamma}) moves from the SEIR-SEI model (Eqs. \ref{modseirHs} - \ref{modseirVi}) to  the delayed model (Eqs. \ref{modseir-retHs} - \ref{modseir-retVi}).

\begin{thebibliography}{99}

\bibitem{amaku2015} {\color{red} Amaku, M., Azevedo, F., Burattini, M. N., Coutinho, F. A. B., Lopez, L. F., \& Massad, E. (2015). Interpretations and pitfalls in modelling vector-transmitted infections. Epidemiology \& Infection, 143(9), 1803-1815.} 

\bibitem{amaku2016} Amaku, M., Azevedo, F., Burattini, M. N., Coelho, G. E., Coutinho, F. A. B., Greenhalgh, D., Lopez, L. F., Motitsuki, R. S., Wilder-Smith, A., \& Massad, E. (2016). Magnitude and frequency variations of vector-borne infection outbreaks using the Ross-Macdonald model: explaining and predicting outbreaks of dengue fever. Epidemiology \& Infection, 144(16), 3435-3450.

\bibitem{anaguano2015}{\color{red} Anaguano, D. F., Ponce, P., Balde\'on, M. E., Santander, S.,
\& Cevallos, V. (2015). Blood-meal identification in phlebotomine sand flies (Diptera: Psychodidae) from Valle Hermoso, a high prevalence zone for cutaneous leishmaniasis in Ecuador. Acta tropica, 152, 116-120.}

\bibitem{aparicio2001}{\color{red} Aparicio, J. P., \& Solari, H. G. (2001). Population dynamics: Poisson approximation and its relation to the langevin process. Physical Review Letters, 86(18), 4183.}

\bibitem{aparicio2007} Aparicio, J. P., \& Pascual, M. (2007). Building epidemiological models from R 0: an implicit treatment of transmission in networks. Proceedings of the Royal Society B: Biological Sciences, 274(1609), 505-512.

\bibitem{auger2008} Auger, P., Kouokam, E., Sallet, G., Tchuente, M., \& Tsanou, B. (2008). The Ross-Macdonald model in a patchy environment. Mathematical biosciences, 216(2), 123-131.

\bibitem{bacaer2006} Baca\"er, N., \& Guernaoui, S. (2006). The epidemic threshold of vector-borne diseases with seasonality. Journal of mathematical biology, 53(3), 421-436.

\bibitem{bacaer2007} Baca\"er, N. (2007). Approximation of the basic reproduction number R0 for vector-borne diseases with a periodic vector population. Bulletin of mathematical biology, 69(3), 1067-1091.

\bibitem{bates2007} Bates, P. A. (2007). Transmission of Leishmania metacyclic promastigotes by phlebotomine sand flies. International journal for parasitology, 37(10), 1097-1106.

\bibitem{belen2006} {\color{red} Belen, A., \& Alten, B. (2006). Variation in life table characteristics among populations of Phlebotomus papatasi at different altitudes. Journal of Vector Ecology, 31(1), 35-44.}

\bibitem{benelli2016} Benelli, G., \& Mehlhorn, H. (2016). Declining malaria, rising of dengue and Zika virus: insights for mosquito vector control. Parasitology research, 115(5), 1747-1754.

\bibitem{bian2004} Bian, L. (2004). A conceptual framework for an individual-based spatially explicit epidemiological model. Environment and Planning B: Planning and Design, 31(3), 381-385.

\bibitem{castanera2003} {\color{red} Casta\~nera, M. B., Aparicio, J. P., \& G\"urtler, R. E. (2003). A stage-structured stochastic model of the population dynamics of Triatoma infestans, the main vector of Chagas disease. Ecological modelling, 162(1-2), 33-53.}



\bibitem{chitnis2008} {\color{red} Chitnis, N., Hyman, J. M., \& Cushing, J. M. (2008). Determining important parameters in the spread of malaria through the sensitivity analysis of a mathematical model. Bulletin of mathematical biology, 70(5), 1272.}

\bibitem{diekman2000} {\color{red} Diekmann, O., \& Heesterbeek, J. A. P. (2000). Mathematical epidemiology of infectious diseases: model building, analysis and interpretation (Vol. 5). John Wiley \& Sons.}

\bibitem{dietz1974}Dietz, K., Molineaux, L., \& Thomas, A. (1974). A malaria model tested in the African savannah. Bulletin of the World Health Organization, 50(3-4), 347.

\bibitem{feng2007} {\color{red} Feng, Z., Xu, D., \& Zhao, H. (2007). Epidemiological models with non-exponentially distributed disease stages and applications to disease control. Bulletin of mathematical biology, 69(5), 1511-1536.}

\bibitem{filigheddu2017} {\color{red} Filigheddu, M. T., G\'orgolas, M., \& Ramos, J. M. (2017). Enfermedad de Chagas de transmisi\'on oral. Medicina cl\'inica, 148(3), 125-131.}


\bibitem{focks2000} {\color{red} Focks, D. A., Brenner, R. J., Hayes, J., \& Daniels, E. (2000). Transmission thresholds for dengue in terms of Aedes aegypti pupae per person with discussion of their utility in source reduction efforts. The American journal of tropical medicine and hygiene, 62(1), 11-18.}

\bibitem{gardner2017} Gardner, L., Chen, N., \& Sarkar, S. (2017). Vector status of Aedes species determines geographical risk of autochthonous Zika virus establishment. PLoS neglected tropical diseases, 11(3), e0005487.

\bibitem{gutierrez2015} Gutierrez, J. A., \& Aparicio, J. P. (2015). Quasi-deterministic population dynamics in stochastic coupled maps. Journal of Biological Systems, 23(supp01), S151-S162.

\bibitem{nepomuceno2016} Nepomuceno, E.G., Takahashi, R. H. C., \& Aguirre, L. A. (2016). Individual-based model (IBM): an alternative framework for epidemiological compartment models. Revista Brasileira de Biometria, 34(1), 133-162.

\bibitem{noireau2009} Noireau, F., Diosque, P., \& Jansen, A. M. (2009). Trypanosoma cruzi: adaptation to its vectors and its hosts. Veterinary research, 40(2), 1-23.

\bibitem{nouvellet2013} {\color{red} Nouvellet, P., Dumonteil, E., \& Gourbière, S. (2013). The improbable transmission of Trypanosoma cruzi to human: the missing link in the dynamics and control of Chagas disease. PLoS neglected tropical diseases, 7(11), e2505.}

\bibitem{macdonald1955} Macdonald, G. (1955). The measurement of malaria transmission. Proceedings of the Royal Society of Medicine, 295-302.

\bibitem{mandal2011} Mandal, S., Sarkar, R. R., \& Sinha, S. (2011). Mathematical models of malaria-a review. Malaria journal, 10(1), 202.

\bibitem{mohammadi2016} {\color{red} Mohammadi, F. (2016). A computational wavelet method for numerical solution of stochastic Volterra-Fredholm integral equations. Wavelet and Linear Algebra, 3(1), 13-25.}

\bibitem{molineaux1980} {\color{red} Molineaux, L., \& Gramiccia, G. (1980). The Garki project: research on the epidemiology and control of malaria in the Sudan savanna of West Africa. World Health Organization.}

\bibitem{oregan2016}O'Regan, S. M., Lillie, J. W., \& Drake, J. M. (2016). Leading indicators of mosquito-borne disease elimination. Theoretical ecology, 9(3), 269-286.

\bibitem{otero2010} {\color{red} Otero, M., \& Solari, H. G. (2010). Stochastic eco-epidemiological model of dengue disease transmission by Aedes aegypti mosquito. Mathematical biosciences, 223(1), 32-46.}

\bibitem{otero2011} Otero, M., Barmak, D. H., Dorso, C. O., Solari, H. G., \& Natiello, M. A. (2011). Modeling dengue outbreaks. Mathematical biosciences, 232(2), 87-95.

\bibitem{rabinovich1972} {\color{red} Rabinovich, J. E. (1972). Vital statistics of Triatominae (Hemiptera: Reduviidae) under laboratory conditions. I. Triatoma infestans Klug. Journal of Medical Entomology, 9(4), 351-370.}


\bibitem{reiner2013} {\color{red} Reiner Jr, R.C., Perkins, T.A., Barker, C.M., Niu, T., Chaves, L.F., Ellis, A.M., George, D.B., Le Menach, A., Pulliam, J.R., Bisanzio, D. and Buckee, C., 2013. A systematic review of mathematical models of mosquito-borne pathogen transmission: 1970 - 2010. Journal of The Royal Society Interface, 10(81), p.20120921.}

\bibitem{romerovivas2005} {\color{red} Romero-Vivas, C. M., \& Falconar, A. K. (2005). Investigation of relationships between Aedes aegypti egg, larvae, pupae, and adult density indices where their main breeding sites were located indoors. Journal of the American Mosquito Control Association, 21(1), 15-22.}

\bibitem{ross1911}Ross, R. (1911). Some Quantitative Studies in Epidemiology. 87, 466 - 467.


\bibitem{sanchez2020}{\color{red} Sanchez , F., \& Calvo, J. G. (2020). Dengue model with early-life stage of vectors and age-structure within host. Revista de Matemática: Teoría y Aplicaciones, 27(1), 157-177.}

\bibitem{schofield1997} Schofield, C. J., \& Dujardin, J. P. (1997). Chagas disease vector control in Central America. Parasitology Today, 13(4), 141-144.

\bibitem{smith2011} {\color{red} Smith, H. (2011). Distributed delay equations and the linear chain trick. In An Introduction to Delay Differential Equations with Applications to the Life Sciences (pp. 119-130). Springer, New York, NY. }

\bibitem{smith2004} Smith, D. L., \& McKenzie, F. E. (2004). Statics and dynamics of malaria infection in Anopheles mosquitoes. Malaria journal, 3(1), 13.

\bibitem{smith2012} Smith, D. L., Battle, K. E., Hay, S. I., Barker, C. M., Scott, T. W., \& McKenzie, F. E. (2012). Ross, Macdonald, and a theory for the dynamics and control of mosquito-transmitted pathogens. PLoS pathogens, 8(4), e1002588.

\bibitem{snow2015} Snow, R. W. (2015). Global malaria eradication and the importance of Plasmodium falciparum epidemiology in Africa. BMC medicine, 13(1), 23.

\bibitem{tomasini2017} {\color{red} Tomasini, N., Ragone, P. G., Gourbi\`ere, S., Aparicio, J. P., \& Diosque, P. (2017). Epidemiological modeling of Trypanosoma cruzi: Low stercorarian transmission and failure of host adaptive immunity explain the frequency of mixed infections in humans. PLoS computational biology, 13(5), e1005532.}

\bibitem{velasco1991} Velasco-Hern\'andez, J. X. (1991). An epidemiological model for the dynamics of Chagas' disease. Biosystems, 26(2), 127-134.

\bibitem{wilder-smith2018} Wilder-Smith, A., \& Massad, E. (2018). Estimating the number of unvaccinated Chinese workers against yellow fever in Angola. BMC infectious diseases, 18(1), 185.

\bibitem{wonham2006} {\color{red} Wonham, M. J., Lewis, M. A., Renclawowicz, J., \& Van den Driessche, P. (2006). Transmission assumptions generate conflicting predictions in host-vector disease models: a case study in West Nile virus. Ecology letters, 9(6), 706-725. }

\bibitem{wonham2008} Wonham, M. J., \& Lewis, M. A. (2008). A comparative analysis of models for West Nile virus. In Mathematical epidemiology (pp. 365-390). Springer, Berlin, Heidelberg.

\bibitem{who2017} World Health Organization. (2017). Vector-borne diseases. Retrieved from: https://www.who.int/news-room/fact-sheets/detail/vector-borne-diseases (Last access: September 25, 2019).

\end{thebibliography}
\end{document}